\documentclass[11pt]{article}
\usepackage[utf8]{inputenc}
\usepackage[english]{babel}
\usepackage{csquotes}
\usepackage{geometry}
\usepackage{float}
\usepackage{subcaption}
\usepackage{graphicx}
\usepackage{amsmath}
\usepackage{amssymb}
\usepackage{booktabs}
\usepackage{setspace}
\usepackage{bm}
\usepackage{tabularx}
\usepackage{multirow}
\usepackage{hyperref}
\usepackage{acronym}
\usepackage{caption}
\usepackage{adjustbox}
\usepackage{tikz}
\usetikzlibrary{fit}
\usepackage{tcolorbox}
\usepackage{authblk}
\usepackage{fix-cm}
\usepackage{xcolor}
\usepackage[normalem]{ulem}   

\DeclareRobustCommand{\del}[1]{%
  {\color{red}%
    \ifmmode
      \text{\sout{$\displaystyle #1$}}
    \else
      \sout{#1}
    \fi
  }%
}
\usepackage[normalem]{ulem} 

\usepackage[style=numeric,sorting=none,backend=biber]{biblatex}
\begin{document}



\title{Causal Effect Estimation with TMLE: Handling Missing Data and Near-Violations of Positivity}
\author[1]{Christoph Wiederkehr \thanks{\href{mailto:Christoph.Wiederkehr@stat.uni-muenchen.de}{Christoph.Wiederkehr@stat.uni-muenchen.de}}}
\author[1]{Christian Heumann \thanks{\href{mailto:christian.heumann@stat.uni-muenchen.de}{christian.heumann@stat.uni-muenchen.de}}}
\author[1,2,3,4]{Michael Schomaker \thanks{\href{mailto:michael.schomaker@stat.uni-muenchen.de}{michael.schomaker@stat.uni-muenchen.de}}}

\affil[1]{\small Department of Statistics, 
  Ludwig-Maximilians University Munich, 
  Munich, Germany}

\affil[2]{Centre for Integrated Data and Epidemiological Research, 
  Cape Town, South Africa}

\affil[3]{Institute of Public Health, Medical Decision Making and  Health Technology Assessment, UMIT - University for Health Sciences, Medical Informatics and Technology, Hall in Tirol, Austria}

\affil[4]{Munich Center for Machine Learning (MCML), 
  Ludwig-Maximilians University Munich, 
  Munich, Germany}
\date{}
\maketitle


\begin{abstract}
We evaluate the performance of targeted maximum likelihood estimation (TMLE) for estimating the average treatment effect in missing data scenarios under varying levels of positivity violations. We employ model- and design-based simulations, with the latter using undersmoothed highly adaptive lasso on the 'WASH Benefits Bangladesh' dataset to mimic real-world complexities. Five missingness-directed acyclic graphs are considered, capturing common missing data mechanisms in epidemiological research, particularly in one-point exposure studies. These mechanisms include also not-at-random missingness in the exposure, outcome, and confounders. We compare eight missing data methods in conjunction with TMLE as the analysis method, distinguishing between non-multiple imputation (non-MI) and multiple imputation (MI) approaches. The MI approaches use both parametric and machine-learning models. 
Results show that non-MI methods, particularly complete cases with TMLE incorporating an outcome-missingness model, exhibit lower bias compared to all other evaluated missing data methods and greater robustness against positivity violations across. In Comparison MI with classification and regression trees (CART) achieve lower root mean squared error, while often maintaining nominal coverage rates. Our findings highlight the trade-offs between bias and coverage, and we recommend using complete cases with TMLE incorporating an outcome-missingness model for bias reduction and MI CART when accurate confidence intervals are the priority.
\end{abstract}


\section{Introduction}

Missing data on multiple variables pose a common challenge in observational studies investigating the effects of treatments on health outcomes. Traditional approaches often rely on the classifications of being missing completely at random (MCAR), at random (MAR) and not at random (MNAR) \cite{refLittle2019, refLittle2024}. However, Mohan and Pearl \cite{refPearl} and Mohan et al. \cite{refMohan2021} introduced missingness-directed acyclic graphs (m-DAGs) as an alternative framework to describe missingness mechanisms. This framework enables a more nuanced assessment of whether unbiased and consistent causal effect estimation is possible \cite{refPearl}, depending on the conditions of recoverability with respect to the chosen missing data method \cite{refMohan2021}. Recoverability, in this context, refers to whether a target parameter, such as a causal effect, can be expressed solely in terms of the distribution of the observed, incomplete data—without requiring knowledge of the unobserved data distribution \cite{refBetancur}. Since no general algorithm exists to determine recoverability in arbitrary settings, identification and estimation strategies must be developed on a case-by-case basis \cite{refHolovchak2024}. The procedure involves defining the target estimand without missing data, constructing an m-DAG to represent relationships between variables and missingness indicators, assessing recoverability under m-DAG assumptions, and selecting an appropriate missing data method \cite{refGuideMDag}. Thus, m-DAGs provide a valuable tool for assessing missing data mechanisms in epidemiological research, particularly in point-exposure studies where missing data are common.\cite{refBetancur}.

In causal inference, the average treatment effect (ATE) is a key target parameter, representing the expected difference in outcomes under treatment versus no treatment \cite{refCausalBook}. Estimation of the ATE from observational data generally relies on the assumptions of consistency, conditional exchangeability, and positivity \cite{refCausalBook, RefTmleBasicRose}. Targeted Maximum Likelihood Estimation (TMLE) is commonly used for estimating the ATE due to its double robustness properties \cite{refSchomakerTMLE, refTMLeReview2023, TargetedLearningBook}. Specifically, TMLE combines both outcome and treatment models, ensuring consistent estimates even if one of these models is misspecified \cite{refTMLeReview2023}. Additionally, TMLE is a plug-in estimator, meaning it directly targets the parameter of interest, while allowing for the incorporation of data-adaptive methods, such as machine learning methods via super learning \cite{RefSL2007}. This flexibility makes TMLE suitable for a wide range of data-generating processes and less reliant on strict parametric assumptions, thus enhancing its applicability \cite{TargetedLearningBook}. Previous research within the m-DAG framework has explored various methods for handling missing data in causal effect estimation, including TMLE with data-adaptive approaches \cite{refDashti}. However, the comparative performance of general missing data methods, such as complete case (CC) analysis and multiple imputation (MI), when used in combination with TMLE, remains uncertain across diverse data settings, particularly in the presence of positivity violations.

Positivity, a crucial assumption in causal inference, requires that within each stratum of the confounders, every individual has a non-zero probability of receiving either exposure condition \cite{refCausalBook}. In observational studies, this assumption is often violated when certain subgroups almost never receive a particular treatment due to socioeconomic, demographic, or clinical factors \cite{RefPosVio2012}. Such violations can severely impact the identifiability of causal effects, leading to high variability in estimates \cite{reftmleR}. Although TMLE can partially mitigate these issues by bounding propensity scores away from extremes, it may introduce bias when positivity violations are severe \cite{RefNearPosVio, reftmleR}. Moreover, MI methods, which lack such a protective procedure, may be inconsistent under severe positivity violations, raising concerns about their applicability even when the MAR assumption holds.

The performance of MI depends on the assumptions made about the data-generating process (DGP) and the missingness mechanism (typically MAR). To ensure valid inference, the imputation model must be 'congenial'—that is, correctly specified and properly aligned with the analysis model \cite{refMeng1994, refMIAlternative}. It is essential to evaluate how MI, in conjunction with TMLE, performs in scenarios where data are MNAR or where positivity violations occur. In particular, comparing MI-based approaches to complete case (CC) analysis and the missing covariate missing indicator (MCMI) approach \cite{refMCMIPaper} is crucial for understanding their relative strengths and limitations in handling missing data. These methods represent a spectrum of strategies commonly used in applied research \cite{refDashti, refPopMisMethods}.

Building upon previous studies \cite{refBetancur, refDashti, refZhang2024}, this paper provides recommendations on the suitability of different missing data methods in conjunction with TMLE in settings both with and without positivity violations. First, we extend existing simulation frameworks by incorporating sophisticated model-based DGPs that capture a broad range of variable types and dependencies, including structured near violations of positivity. To systematically evaluate their impact, we vary the degree of near positivity violations and assess how different levels influence estimation.
Second, we bridge the gap between theoretical simulations and practical applications by incorporating a design-based DGP approach. While model-based simulations are widely used, their findings may not always generalize to empirical data, which often deviate from idealized distributions \cite{refGuidanceImpMeth}. To address this limitation, we complement model-based simulations with a data-driven approach that employs undersmoothed highly adaptive lasso \cite{refUnderHAL} on empirical data from the 'WASH Benefits Bangladesh' study \cite{refWashData}. 
Lastly, we expand the comparative analysis of missing data methods by systematically evaluating multiple approaches, both MI and non-MI, when used in conjunction with TMLE. This comprehensive comparison offers new insights into the robustness and limitations of these methods in the presence of positivity violations and complex missing data mechanisms.

\section{Framework}
In this work, we follow the approach proposed by Lee et al. \cite{refGuideMDag}, using m-DAGs to depict missingness assumptions and assess their implications for the recoverability of the causal estimand.

\subsection{Causal Estimand} 
\label{sec:Causal Estimand}
We are interested in estimating the ATE of a binary exposure \( A \) on a continuous outcome \( Y \). By using the potential outcomes framework \cite{refPotentialOut}, we define \( Y(1) \) as the potential outcome that would be observed for an individual if they were exposed (\( A = 1 \)), and \( Y(0) \) as the potential outcome if they were not exposed (\( A = 0 \)). The ATE is the difference in the expected potential outcomes across the population: \(\text{ATE} = E[Y(1)] - E[Y(0)] \).
To identify the ATE from observed data in the absence of missing data, we rely on the following assumptions \cite{TargetedLearningBook, refTmleNearPos, RefTmleBasicRose}: 
\begin{itemize}
    \item Stable Unit Treatment Value Assumption (SUTVA): (i) \emph{consistency}, meaning that if \( A=a \), then the observed outcome equals the corresponding potential outcome, \( Y = Y(a) \); and (ii) \emph{no interference}, so that one individual’s treatment assignment does not affect another individual’s outcome.
    \item Conditional exchangeability (unconfoundedness): the potential outcomes are independent of treatment assignment given measured covariates, i.e. 
    \[
    (Y(1), Y(0)) \perp\!\!\!\perp A \mid W.
    \]
    \item Positivity (overlap): every individual has a positive probability of receiving each treatment level, conditional on their covariates, formally
    \[
    0 < P(A=a \mid W=w) < 1 \quad \forall \, a \in \{0,1\}, \; \forall \, w \in \mathcal{W} \text{ with } P(W=w) > 0.
    \]
    In practice, positivity violations arise when the estimated propensity scores \( \hat{P}(A = a \mid \boldsymbol{W} = w) \) approach 0 or 1 \cite{RefPosVio2012}.
\end{itemize}. 
Under these assumptions, the ATE can be identified and expressed in terms of the observed data as follows \cite{refCausalBook}:
\[
\text{ATE} = \sum_{w \in \mathcal{W}} \left( E[Y \mid \boldsymbol{W} = w, A = 1] P(\boldsymbol{W} = w) \right) - \sum_{w \in \mathcal{W}} \left( E[Y \mid \boldsymbol{W} = w, A = 0] P(\boldsymbol{W} = w) \right)
\]
Here, \( E[Y \mid \boldsymbol{W} = w, A = a] \) represents the expected value of the outcome \( Y \) given the vector of confounders \( \boldsymbol{W} = w \) and the exposure level \( A = a \), while \( P(\boldsymbol{W} = w) \) denotes the probability of the confounder vector \( \boldsymbol{W} \) taking the realization \( w \).

\subsection{Estimation: Targeted Maximum Likelihood Estimation (TMLE)}
\label{sec:TMLE}

TMLE \cite{TargetedLearningBook} is a semiparametric estimation method that improves efficiency and reduces bias in causal effect estimation. It begins by fitting an initial model to estimate the conditional expectation of the continuous outcome \( Y \), given the binary treatment \( A \) and the vector of confounders \( \boldsymbol{W} \):
\( \hat{Q}(A, \boldsymbol{W}) = \hat{E}[Y \mid A, \boldsymbol{W}] \). In practice, Y is often transformed to lie within the range [0,1], to facilitate modeling with logistic regression.
Next,  the propensity score \( \hat{g}(A \mid \boldsymbol{W}) \) is estimated, representing the probability of receiving the treatment given the confounders:
\( P(A = 1 \mid \boldsymbol{W}) \).
To improve bias reduction, TMLE constructs clever covariates, which serve as inverse probability weights in the updating step:
\[
\hat{H}(A = a, \boldsymbol{W}) = \frac{\mathbb{I}\{A = a\}}{\hat{g}( A= a \mid \boldsymbol{W})} \qquad (a \in \{0,1\})
\]
To incorporate information from the treatment mechanism and improve efficiency, the initial estimate \( \hat{Q}(A, \boldsymbol{W}) \) is refined through a fluctuation model. Specifically, a logistic regression is performed, where the logit of the initial estimate serves as an offset, and the clever covariates \( \hat{H}(1, \boldsymbol{W}) \) and \( \hat{H}(0, \boldsymbol{W}) \) are used as predictors. The estimated fluctuation parameter \( \hat{\epsilon} \) adjusts the initial prediction:
\[
\hat{Q}^*(A, \boldsymbol{W}) = \text{logit}^{-1}\left(\text{logit}(\hat{Q}(A, \boldsymbol{W})) + \hat{\epsilon} \cdot \hat{H}(A, \boldsymbol{W})\right)
\]
This targeted update \( \hat{Q}^*(A, \boldsymbol{W}) \) ensures that the final outcome estimate properly incorporates information from the estimated propensity score, thereby reducing bias and improving efficiency.
The ATE is then estimated by averaging the differences in the updated (targeted) outcome predictions across all individuals:
\[
\hat{\text{ATE}} = \frac{1}{n} \sum_{i=1}^n \left( \hat{Q}^*(1, \boldsymbol{W}_i) - \hat{Q}^*(0, \boldsymbol{W}_i) \right)
\]
TMLE is double robust, meaning it remains consistent as long as either the outcome model or the treatment model is correctly specified \cite{refSchomakerTMLE}. To improve robustness against model misspecification, we use SuperLearner, an ensemble learning method that combines multiple base learners through cross-validation \cite{RefSL2007}. In our implementation, we include generalized linear models (with and without interactions), multivariate adaptive regression splines, and the arithmetic mean as base learners. SuperLearner selects an optimal weighted combination of these models, reducing the risk of misspecification and improving predictive performance. The estimation is performed using the \texttt{SuperLearner} package in R \cite{refSLR}. To address extreme propensity scores, we apply truncation, setting estimated propensity scores below 0.01 to 0.01. This helps stabilize estimates and prevents excessive influence from sparsely represented subgroups. Lastly, the variance of the ATE is estimated using the efficient influence curve-based variance estimator \cite{TargetedLearningBook}.

\subsection{Missingness Mechanisms: m-DAGs}
\begin{figure}[htp]
  \centering
  \caption[m-DAGs]{\fontsize{9}{11}\selectfont Missingness directed acyclic graphs (m-DAGs) illustrating the five missingness scenarios considered in the model-based simulation study. Figure adapted from Moreno-Betancur et al.~\cite{refBetancur}.}
  \includegraphics[scale=0.92]{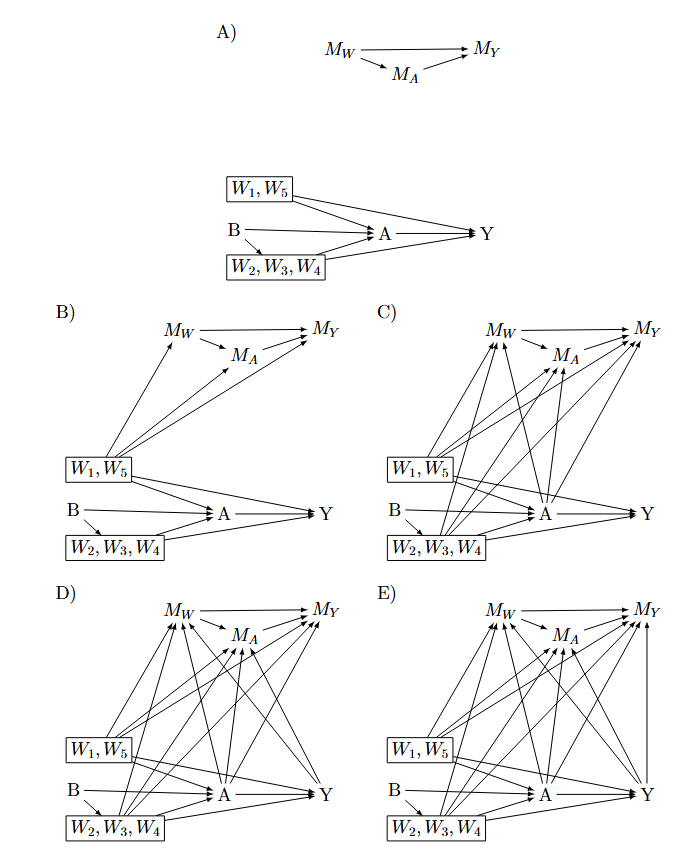} 
\caption*{\fontsize{9}{11}\selectfont Each node corresponds to a study variable: an auxiliary variable \(B\), confounders \((W_1,W_5)\) (fully observed) and \((W_2,W_3,W_4)\) (partially observed), the exposure \(A\), and the outcome \(Y\). 
The missingness indicators are denoted by \(M_A\) (exposure), \(M_Y\) (outcome), and \(M_W\) (partially observed confounders). 
Arrows into a missingness node indicate which variables influence whether the corresponding variable is observed. 
For example, \(V\!\to\!M_V\) represents self-masked missingness and e.g. \(Y\!\to\!M_A\) represents exposure recording depending on the outcome.
The five m-DAGs capture increasingly complex and realistic missingness scenarios. 
m-DAG A (the trivial m-DAG) corresponds to MCAR, where no arrows point into any missingness node, so missingness is independent of all modeled variables. 
m-DAG B represents MAR, where missingness depends only on fully observed variables \((B,W_1,W_5)\) and not on the unobserved values themselves. 
m-DAGs C–E represent MNAR mechanisms, in which missingness depends on partially observed or unobserved variables.} 
\label{fig:m-DAGs}
\end{figure}

We consider five m-DAGs, adapted from Moreno-Betancur et al.~\cite{refBetancur}; see Figure~\ref{fig:m-DAGs}. M-DAGs extend standard DAGs by explicitly modeling missing data mechanisms. In addition to capturing causal relationships between observed variables, m-DAGs include missingness indicators that encode assumptions about the processes leading to missing data. Arrows from observed variables to missingness indicators represent dependencies, indicating that the probability of missingness may be influenced by specific variables.
The set of confounders \(\mathbf{W}\) is partitioned into fully observed confounders \((W_1,W_5)\) and incompletely observed confounders \((W_2,W_3,W_4)\). 
For any possibly incomplete variable \(V\in\{A,Y,W_2,W_3,W_4\}\) with missingness indicator \(M_V\in\{0,1\}\), we denote the recorded value by
\[
V^{\mathrm{obs}}=
\begin{cases}
V, & M_V=0,\\
\text{missing}, & M_V=1.
\end{cases}
\]
Thus, the available data consist of \((B, W_1,W_5)\) (fully observed), \(\{V^{\mathrm{obs}}:V\in\{A,Y,W_2,W_3,W_4\}\}\), and the missingness indicators \(M_A,M_Y,M_W\), where \(M_W\) collects the indicators for \(W_2,W_3,W_4\).
Arrows into an \(M\)-node encode which variables affect the probability that the corresponding \(V\) is observed: \(V\!\to\!M_V\) represents self-masked missingness of \(V\); \(A\!\to\!M_Y\) means observation of \(Y\) depends on exposure ; \(Y\!\to\!M_A\) means recording of \(A\) depends on the outcome. 
Conceptually, such arrows indicate situations where missingness is informative: for instance, if individuals with worse outcomes are less likely to have their exposure measured, then \(Y\!\to\!M_A\).
These m-DAGs in figure \ref{fig:m-DAGs}, with the exception of m-DAG A, represent distinct common missingness scenarios in epidemiological point-exposure studies \cite{refBetancur}. Using an alternative framework of missing data taxonomy based on graphical models \cite{refPearl}, these mechanisms can be classified as follows:

\begin{itemize}
    \item m-DAG A (MCAR): no arrows from \(A,Y,\mathbf{W}\) into any \(M\)-node; missingness is independent of all modeled variables.
    \item m-DAG B (MAR): each \(M\)-node depends only on the fully observed variables \((B,W_1,W_5)\), not on the missing values themselves.
    \item m-DAGs C, D and E (MNAR): at least one incomplete variable directly influences its own missingness (\(V\!\to\!M_V\)) and/or \(M\)-nodes depend on other incomplete variables; hence MCAR/MAR do not hold. For example, \(Y\!\to\!M_Y\) corresponds to outcome-dependent follow-up, where patients with worse outcomes are less likely to have the outcome recorded.
\end{itemize}

Following Moreno-Betancur et al. \cite{refBetancur}, we adopt their structures but, for simulation control of joint missingness, we allow direct arrows among the missingness indicators (e.g., \(M_A\!\to\!M_Y\)), rather than representing such associations solely via an unmeasured common cause.

\subsection{Recoverability of the estimand in the respective m-DAGs}
Recoverability refers to the ability to consistently estimate a target parameter, such as the ATE, from the available data, even in the presence of missing data. We proceed in two stages. First, in a hypothetical setting without missing data, we invoke the standard causal identification assumptions—consistency, no interference, conditional exchangeability given $\boldsymbol{W}$, and positivity - under which the ATE can be expressed through the g-formula (see Section \ref{sec:Causal Estimand}).

Second, given that the estimand is identified in complete data, we consider the observed-data world with missingness as encoded by an m-DAG and ask whether this identified functional is recoverable from the observed-data distribution. By recoverability we mean that the g-formula can be re-expressed purely in terms of observables such that any variable subject to missingness appears within probability statements conditional on it being observed \cite{refPearl}. This ensures that missingness does not preclude consistent estimation of the identified ATE. The critical quantities that need to be expressible are \( E \left( Y \mid A, W_{2,3,4}, W_1, W_5 \right) \) and \( P(W_{2,3,4}) \). The detailed derivation of recoverability (similar to Moreno-Betancur et al. \cite{refBetancur}) for the specific m-DAGs considered in this study can be found in Appendix \ref{Appendix_Recov}.
In brief, the ATE is recoverable for m-DAGs A, B and C, given the standard causal identification assumptions are met. However the ATE is not recoverable in m-DAGs D and E for any available case method (Table \ref{tab:Recov}). 

\subsection{Handling Missing Data}
\label{sec:Handling Missing Data}
Various approaches to handle missing data were used, including both non-MI and MI methods. Table \ref{tab:Recov} summarizes the recoverability of the ATE using the described missing data methods across all considered m-DAGs.

\subsubsection{Non-MI Methods}

\begin{itemize}
    \item \textbf{Complete-Case Analysis (CC)}: This approach restricts the analysis to observations with no missing data in any of the variables used for estimation. Thus TMLE is performed on the subset of data where all confounders, treatment, and outcome variables are fully observed.
    
    \item \textbf{Extended TMLE (Ext)}: This approach discards observations with missing values in confounders and exposure variables. Then a model for outcome missingness is incorporated into the targeting step of TMLE (Appendix \ref{Append:Unbiased estimation for different missing data methods}).
    
    \item \textbf{Extended TMLE with Missing Covariate Missing Indicator (Ext MCMI)}: Only observations with missing exposure data are excluded, and missingness indicators are included for incomplete confounders \cite{refMCMIPaper}. Then the extended TMLE approach is used to handle outcome data \cite{refDashti}.
\end{itemize}

\subsubsection{Fully Conditional Specification (FCS) Methods}

FCS imputations methods, i.e. chained equation approaches, were implemented using the \texttt{mice} package in R \cite{refmiceR}, which iteratively imputes missing values by sampling from a series of univariate conditional models. We performed 100 imputations, following the rule of thumb proposed by White et al. \cite{refMIAlternative} to achieve sufficient accuracy for method comparison. Oberman et al. \cite{refGuidanceImpMeth} determined that 5-10 iterations are generally adequate for the convergence of MI algorithms, even under conditions of substantial missingness. Therefore, we set the number of iterations to 10. The results of the analysis on the imputed data are pooled using Rubin’s rules to obtain the final estimate and its standard errors \cite{refStefBuurenBook}. All available variables, including the outcome, are used for imputation, as recommended by prior research \cite{refWhyYinImp}.

\begin{itemize}
    \item \textbf{MI using Predictive Mean Matching (MI PMM)}: This method uses parametric imputation using predictive mean matching for the outcome Y and applies appropriate models for other variables based on their type, including logistic regression for binary variables, polytomous regression for categorical variables, and Bayesian linear regression for continuous variables \cite{refmiceR}.
    
    \item \textbf{MI with Interaction Terms (MI Int)}: Extends MI PMM by appropriately incorporating two-, three-, and four-way interactions into the imputation models \cite{refAppropriateInteractions}.
    
    \item \textbf{MI using Classification and Regression Trees (MI CART)}: Non-parametric approach utilizes decision trees for all imputation models.
    
    \item \textbf{MI using Random Forest (MI RF)}: This method uses random forests for all imputation models.
\end{itemize}

\subsubsection{Joint Multivariate Normal MI}

The \textbf{MI using Amelia} approach assumes that the hypothetically complete data follow a multivariate normal distribution, possibly after suitable transformations beforehand. This joint modeling (JM) method is implemented using the \texttt{Amelia} package in R \cite{refAmeliaR}, using the EMB-algorithm \cite{EMBAlgo}. \texttt{Amelia} automatically iterates until convergence is achieved, eliminating the need for a predefined number of iterations. Although the normality assumption may seem restrictive, it has been shown to often perform well, even with categorical or mixed data types \cite{refMIAmeliaSufficient}, \cite{refAmeliaR}.

\begin{table}[H]
\centering
\caption{\small Indicates whether the ATE (average treatment effect) is recoverable in the given m-DAG (missingness-directed acyclic graph). “Yes” indicates that unbiased estimation of the ATE is possible using the respective missing data method in conjunction with TMLE (targeted maximum likelihood estimation), provided that the standard causal identification assumptions hold. Note, however, that in scenarios with positivity violations (Levels 2 and 3), these assumptions are not fully satisfied, and the ATE cannot be estimated consistently. Further details are provided in Appendix \ref{Appendix_Recov}.}
\label{tab:recoverability}
\begin{tabular}{lccccc}
\toprule
\textbf{m-DAG} & \textbf{ATE} & \textbf{CC} & \textbf{Ext} & \textbf{Ext MCMI} & $\text{\textbf{MI}}^{b}$ \\
\midrule
\textbf{A}     & Recoverable & Yes & Yes & No & Yes \\
\textbf{B}     & Recoverable & Yes & Yes & No & Yes \\
\textbf{C}     & $\text{  Recoverable}^{a}$ & $\text{  Yes}^{a}$ & $\text{  Yes}^{a}$ & No & No \\
\textbf{D}     & Non-recoverable & No & No & No & No \\
\textbf{E}     & Non-recoverable & No & No & No & No \\
\bottomrule
\end{tabular}

\caption*{\small
Abbreviations:  
CC = complete case analysis with TMLE;  
Ext = extended TMLE incorporating an outcome-missingness model;  
Ext MCMI = extended TMLE plus missing covariate missing indicator approach;  
MI = multiple imputation with TMLE as analysis method. \\  
$^{a}$ Holds only in model-based simulation, as no treatment effect heterogeneity (i.e., effect modification) was introduced. In the design-based simulation, the undersmoothed highly adaptive lasso (HAL) fit for the conditional outcome included few basis functions representing confounder–exposure interactions. \\
$^{b}$ MI provides merely consistent ATE estimation for m-DAG versions of MCAR (missing completely at random) or MAR (missing at random) \cite{refMohan2021}.}

\label{tab:Recov}
\end{table}

\begin{table}[H]
\centering
\scriptsize
\caption{Overview of the key features of the simulation studies}
\begin{tabular}{|p{2.5cm}|p{14cm}|}
\hline
\textbf{Key features of the simulation study} & \multicolumn{1}{c|}{\textbf{Description}} \\ \hline
Aim & Evaluate the performance of general missing data methods combined with Targeted Maximum Likelihood Estimation (TMLE) for estimating causal effects for varying levels of positivity violations. \\ \hline
Data generating mechanism & 
Model-based Simulation (see Section~\ref{sec:DGP_ModelBased}):
\begin{itemize}
    \item Incorporate a range of \underline{different variable types} for each DGP (1-5).
    \item \underline{Three positivity levels (scenarios)} were examined across all five DGPs: 
          the exposure variable was generated from regression models with 
          \underline{varying interaction effect sizes}, leading to differing degrees of 
          positivity violation.
    \item Amount of positivity violations: Level 1 $\leq 1\%$; Level 2 $\sim 10\%$; Level 3 $\sim 30\%$.
\end{itemize}
Design-based Simulation on 'WASH Benefits Bangladesh' data (see Section~\ref{sec:DGP_DesginBased}):
\begin{itemize}
    \item  Estimate conditional distribution of Exposure and Outcome with the \underline{undersmoothed Highly Adaptive Lasso}
    \item Draw bootstrap samples and use the undersmoothed Highly Adaptive Lasso fit for generating synthetic data
    \item No positivity violation
\end{itemize} \\ \hline
Missingness mechanism & 
Consider \underline{five} missing Directed Acyclic Graphs (m-DAGs) (see Figure~\ref{fig:m-DAGs}):
\begin{itemize}
    \item One m-DAG serves as a benchmark, representing missingness completely at random (MCAR).
    \item One m-DAG serves as a benchmark for MI-methods, representing missingness at random (MAR).
    \item Three m-DAGs represent common missing data mechanisms encountered in epidemiological research and fall under the category of missing not at random (MNAR). Among those, only one allows for the consistent estimation of the causal effect using a complete case analysis.
\end{itemize} \\ \hline
Missing data methods & Compare several methods (see Section~\ref{sec:Handling Missing Data}):
\begin{itemize}
    \item Complete Cases (CC) analysis.
    \item CC with TMLE incorporating outcome-missingness models (Ext).
    \item Missing covariate indicator method in conjunction with TMLE incorporating outcome-missingness models (Ext MCMI).
    \item Four fully conditional specification (FCS) approaches, also known as Multiple Imputation using Chained Equations (MICE), employing both parametric and machine-learning models.
    \item Joint multivariate normal MI method with the EMB-Algorithm.
\end{itemize} \\ \hline
Method of analysis & TMLE using a data adaptive approach via the Super Learner (see Section~\ref{sec:TMLE}). \\ \hline
Performance measures &  (see Section~\ref{sec:Evaluation criteria})
\begin{itemize}
    \item Assess relative bias.
    \item Assess the root mean square error (RMSE).
    \item Assess the nominal coverage.
\end{itemize} \\ \hline
Number of repetitions & 1000 (Figure~\ref{fig:Flowchart} illustrates the workflow for the model-based simulation) \\ \hline
\end{tabular}
\normalsize
 \label{tab:Keyfeatures_simulationstudy}
\end{table}

\section{Model-based Simulations}

We implemented five distinct DGPs, each with three scenarios varying in positivity violation severity (see Table \ref{tab:Keyfeatures_simulationstudy}). The DGPs were constructed sequentially, with each extending the previous one (e.g., DGP 2 builds on DGP 1). For each scenario within each DGP, we generated 1,000 datasets, each containing 2,000 observations.

\subsection{Data generating processes}
\label{sec:DGP_ModelBased}
DGP1 follows the simulation design of Dashti et al. \cite{refDashti}, which was constructed to approximate the structure of the Victorian Adolescent Health Cohort Study (VAHCS) \cite{refVAHCS}. For all scenarios in DGP 1, an auxiliary variable B with a standard normal distribution was generated, along with a set of binary confounders \( \boldsymbol{W}=\left(W_{1}, W_{2}, W_{3}, W_{4}, W_{5}\right) \). Confounders \( W_{2}, W_{3} \), and \( W_{4} \) were generated through regression on  B: 
\begin{align*}
\mathrm{B} &\sim \mathcal{N}(0,1) \\
\mathrm{W}{1} &\sim \operatorname{Binomial}\left(1, \operatorname{logit}^{-1}\left(\alpha_{0}\right)\right) \\
\mathrm{W}{2} &\sim \operatorname{Binomial}\left(1, \operatorname{logit}^{-1}\left(\beta_{0}+\beta_{1} \mathrm{B}\right)\right) \\
\mathrm{W}{3} &\sim \operatorname{Binomial}\left(1, \operatorname{logit}^{-1}\left(\gamma_{0}+\gamma_{1} \mathrm{B}\right)\right) \\
\mathrm{W}{4} &\sim \operatorname{Binomial}\left(1, \operatorname{logit}^{-1}\left(\delta_{0}+\delta_{1} \mathrm{B}\right)\right) \\
\mathrm{W}{5} &\sim \operatorname{Binomial}\left(1, \operatorname{logit}^{-1}\left(\zeta_{0}\right)\right),
\end{align*}
\noindent where \( \operatorname{logit}^{-1}(\cdot)=\exp (\cdot) /(1+\exp (\cdot)). \) Exposure \( \mathrm{A} \) was modeled via regression on \( \mathrm{B}, \mathrm{W} \) incorporating two-way confounder-confounder interactions. In the baseline positivity scenario, we made minor adjustments to the interaction coefficients compared to Dashti et al. \cite{refDashti} to better control the degree of positivity. 
To generate increasing degrees of positivity violation, the interaction coefficients were progressively inflated: 
at each higher positivity level, the coefficients were approximately doubled compared to the previous level. 
This exaggeration of interactions creates subgroups of \( W\) in which the conditional probability of exposure 
\( \mathrm{A} \) becomes very small, thereby inducing greater sparsity while maintaining an overall exposure prevalence of 15\%. The outcome \( \mathrm{Y} \) was normally distributed with a mean of 0 and a standard deviation of 1, whereas the mean was generated through regression on \( \mathrm{A} \) and \( \boldsymbol{W} \) involving two-, three-, and four-way confounder-confounder interactions. However, the interaction effect sizes remained constant across all three scenarios:
\begin{align*}
\mathrm{A} &\sim \operatorname{Binomial}\left(1, \operatorname{logit}^{-1}\left(\eta_{0}+\eta_{1} \mathrm{W}_1 +\eta_{2} \mathrm{W}_{2}+\eta_{3} \mathrm{W}_{3}+\eta_{4} \mathrm{W}_{4}+\eta_{5} \mathrm{W}_{5}+\eta_{6} \mathrm{B} +\eta_{7} \mathrm{W}_{1} \mathrm{W}_{3}\right.\right. \\ &\quad\left.\left.+\eta_{8} \mathrm{W}_{1} \mathrm{W}_{4} +\eta_{9} \mathrm{W}_{1} \mathrm{W}_{5}+\eta_{10} \mathrm{W}_{3} \mathrm{W}_{4}+\eta_{11} \mathrm{W}_{3} \mathrm{W}_{5}+\eta_{12} \mathrm{W}_{4} \mathrm{W}_{5}\right)\right) \\ \mathrm{Y} &\sim \mathcal{N}\left(\upsilon_{0}+\upsilon_{1} \mathrm{A}+\upsilon_{2} \mathrm{W}_{1}+\upsilon_{3} \mathrm{W}_{2}+\upsilon_{4} \mathrm{W}_{3}+\upsilon_{5} \mathrm{W}_{4}+\upsilon_{6} \mathrm{W}_{5}+\upsilon_{7} \mathrm{W}_{1} \mathrm{W}_{3}+\upsilon_{8} \mathrm{W}_{1} \mathrm{W}_{4}+\upsilon_{9} \mathrm{W}_{1} \mathrm{W}_{5}\right. \\ &\quad+\upsilon_{10} \mathrm{W}_{3} \mathrm{W}_{4} +\upsilon_{11} \mathrm{W}_{3} \mathrm{W}_{5}+\upsilon_{12} \mathrm{W}_{4} \mathrm{W}_{5}+\upsilon_{13} \mathrm{W}_{1} \mathrm{W}_{3} \mathrm{W}_{4}+\upsilon_{14} \mathrm{W}_{1} \mathrm{W}_{3} \mathrm{W}_{5} \\ &\quad\left.+\upsilon_{15} \mathrm{W}_{1} \mathrm{W}_{4} \mathrm{W}_{5}+\upsilon_{16} \mathrm{W}_{3} \mathrm{W}_{4} \mathrm{W}_{5} +\upsilon_{17} \mathrm{W}_{1} \mathrm{W}_{3} \mathrm{W}_{4} \mathrm{W}_{5}, \sigma=1\right)
\end{align*}
\noindent Across all outcome generation models, the coefficient for \(\mathrm{A}\left( \upsilon_{1}\right) \), representing the true ATE, varied between 0.18 and 0.24 across all DGPs and scenarios. This variation was chosen to ensure 80\% statistical power for detecting the effect across the 1,000 simulated datasets. For DGP 2 both binary confounders \(  W_{4}, W_{5} \) were replaced with continuous variables, drawn from a normal distribution. The mean of \(  W_{4} \) was generated via regression on \( B \):
\begin{align*}
\mathrm{W}{4} &\sim \mathcal{N} \left(\delta_{0}+\delta_{1} \mathrm{B}, \sigma = 1 \right) \\
\mathrm{W}{5} &\sim \mathcal{N} \left( \zeta_{0}, \sigma = 2 \right).
\end{align*}
The extension from DGP 2 to DGP 3 involves utilizing the \texttt{copula} r-package \cite{refcopulaR} to include interdependencies between the different confounders \( \boldsymbol{W}=\left(W_{1}, W_{2}, W_{3}, W_{4}, W_{5}\right) \) with a Gaussian copula. In DGP 4 the binary confounder \( W_3 \) is replaced by a categorical variable with four categories. The probabilities of each category depend on \(B \) and are created using a softmax function:
\[
P(W_3 = i) = \frac{e^{\gamma_{0_i} + \gamma_{1_i} \cdot B}}{\sum_{j=1}^{4} e^{\gamma_{0_j} + \gamma_{1_j} \cdot B}}, \quad \text{for } i,j = 1, ..., 4.
\]
Lastly in DGP 5 a gamma-distributed confounder \( W_6 \) is included (see Appendix \ref{Append:ModelBasedDGP}). The shape and rate parameters of the gamma distribution are linear functions of \( t_B \), the truncated version of \( B \) with the range of \( [-0.99, 0.99] \). This requirement ensures that the parameters of the gamma distribution remain positive.
The equations for the parameters are \( \text{shape} = \xi_0 + \xi_1t_B \) and \( \text{rate} = \tau_0 + \tau_1t_B \).
Across all five DGPs, the positivity‐violation rate was \(\leq 1\%\) at positivity level 1, \(\sim 10\%\) at level 2, and \(\sim 30\%\) at level 3.

To quantify the extent of positivity violations in the data, the following procedure was used on the full data before inducing missingness:
\begin{enumerate}
    \item \textbf{Propensity Score Estimation:} The same logistic regression model, which includes interaction terms among the confounders \( \boldsymbol{W} \), is re-estimated on the generated data to obtain estimated propensity scores \( \hat{P}(A = 1 \mid \mathbf{W}) \)
    \item \textbf{Identification of Violations:} For each observation \( i \), compute the estimated propensity score \( \hat{p}_i = \hat{P}(A = 1 \mid \mathbf{W}_i) \). An observation is considered to exhibit a positivity violation if:
    \[
        \min(\hat{p}_i, 1 - \hat{p}_i) < 0.002
    \]
    \item \textbf{Quantification of Violations:} Calculate the proportion of observations that satisfy the above condition.
\end{enumerate}
The logistic regression model, incorporating interaction terms, is assumed to accurately represent the true DGP for the exposure \( A \). Consequently, re-estimating this model on the generated data yields precise estimates of the propensity scores. \newline
Further details on the variable distributions and guassian copula are in Appendix \ref{Append:ModelBasedDGP}.

\subsection{Generating missingness}
\label{sec:model_generatingMissingness}  Missingness was imposed on \( W_{2}, W_{3}, W_{4}, A \) and \( Y \) by generating missingness indicators \( M_{W_2}, M_{W_3}, M_{W_4}, M_A \) and \( M_{Y} \), which were coded 1 if the variable was missing and 0 if observed. The models used for generating these missingness indicators in the DGP1, DGP2, DGP3 and DGP4 were as follows:
\begin{align*} 
\mathrm{M}_{\mathrm{W}_{2}} &\sim \operatorname{Binomial}(1, \operatorname{logit}^{-1}(\iota_{0} + \iota_{1} \mathrm{W}_{1} + \iota_{2} \mathrm{W}_{5} + \iota_{3} \mathrm{W}_{2}+\iota_{4} A + \iota_{5} \mathrm{Y}) ) \\ 
\mathrm{M}_{\mathrm{W}_{3}} & \sim \operatorname{Binomial}(1, \operatorname{logit}^{-1} ( \kappa_{0} + \kappa_{1} \mathrm{W}_{1} + \kappa_{2} \mathrm{W}_{5} + \kappa_{3} \mathrm{W}_{3} + \kappa_{4} A + \kappa_{5} \mathrm{Y} + \kappa_{6} \mathrm{M}_{\mathrm{W}_{2}} ) ) \\ 
\mathrm{M}_{\mathrm{W}_{4}} &\sim \operatorname{Binomial} (1, \operatorname{logit}^{-1} ( \lambda_{0} + \lambda_{1} \mathrm{W}_{1} + \lambda_{2} \mathrm{W}_{5} + \lambda_{3} \mathrm{W}_{4} + \lambda_{4} A + \lambda_{5} \mathrm{Y} + \lambda_{6} \mathrm{M}_{\mathrm{W}_{2}} + \lambda_{7} \mathrm{M}_{\mathrm{W}_{3}} ) ) \\
\mathrm{M}_{A} &\sim \operatorname{Binomial} (1, \operatorname{logit}^{-1} ( \nu_{0}+ \nu_{1} \mathrm{W}_{1} + \nu_{2} \mathrm{W}_{5}+  \nu_{3} \mathrm{W}_{2} + \nu_{4} \mathrm{W}_{3} + \nu_{5} \mathrm{W}_{4} + \nu_{6} A + \nu_{7} \mathrm{Y} \\ &\quad + \nu_{8} \mathrm{M}_{\mathrm{W}_{2}} + \nu_{9} \mathrm{M}_{\mathrm{W}_{3}} + \nu_{10} \mathrm{M}_{\mathrm{W}_{4}} ) ) \\
\mathrm{M}_{\mathrm{Y}} &\sim \operatorname{Binomial} (1 , \operatorname{logit}^{-1} (\xi_{0}+\xi_{1} \mathrm{W}_{1}+\xi_{2} \mathrm{W}_{5}+ \xi_{3} \mathrm{W}_{2}+\xi_{4} \mathrm{W}_{3}+\xi_{5} \mathrm{W}_{4}+\xi_{6} A +\xi_{7} \mathrm{Y}+\xi_{8} \mathrm{M}_{\mathrm{W}_{2}} \\ &\quad + \xi_{9} \mathrm{M}_{\mathrm{W}_{3}} +\xi_{10} \mathrm{M}_{\mathrm{W}_{4}}+\xi_{11} \mathrm{M}_{A} ) )
\end{align*}
\noindent When an arrow was present in the considered m-DAG, the coefficients for binary confounders were set to 0.6. For continuous confounders and the outcome, a similar approach was employed, with the coefficient set to 0.2.  
In these missingness models, we assign negative signs to the coefficients of $A$ and $W_5$, while all other covariate coefficients are positive. This strategy was intended to balance the influence of different variable types. The models used for generating the missingness indicators undergo minor adjustments when \( W_6 \) is incorporated in DGP 5. The missingness proportions accross the scenarios and m-DAGs remain the same and are depicted in Table \ref{tab:missingness_proportion}. The model equations for DGP5 are provided in the Appendix \ref{Append:ModelBasedMissing}.

\begin{table}[H]
\centering
\fontsize{10}{12}\selectfont
\begin{adjustbox}{max width=\textwidth, max height=\textheight}
\scalebox{1}{
\begin{tabular}{lccccccccccc}
 \hline
  & \multicolumn{8}{c}{\textbf{\% with missing value}}  &  &  &  \\
   & W2 & W3 & W4 & W6 & A & Y & A/Y & Any &  &  &  \\ \cline{1-10}
\multirow{1}{*}{For all scenarios} and all m-DAGs & 25 & 30 & 25 & 30 & 30 & 20 & 40 & 50 &  &  &  \\
\end{tabular}
}
 \end{adjustbox}
 \caption[Distribution and missingness of variables]{\small Proportion of missingness by variable across the five missingness mechanisms (m-DAGs) and all scenarios.
 “A/Y” denotes the proportion with a missing value in either $A$ or $Y$; “Any” denotes the proportion with a missing value in any of $A$, $Y$, or $W_{2,3,4}$. 
By design, these margins are held constant across m-DAGs and scenarios. DGPs~1--4 do not include $W_6$.}
    \label{tab:missingness_proportion}
\end{table}

In the study, we explicitly incorporated missing mechanisms through m-DAGs to address concerns raised by Schouten et al. \cite{refSimMissingness} and ensure that the implemented missingness matched the intended patterns. We varied coefficient signs to keep the linear predictor for the respective missingness model balanced and to avoid saturation of the logistic link (where a large intercept would make predicted probabilities insensitive to covariates). Continuous predictors were standardized, and their coefficients ($0.2$) were set such that an increase from one standard deviation below to one standard deviation above the mean corresponds to a change of approximately $0.4$ in the log-odds.
For each missingness indicator $M$, the intercept was then chosen to match the target marginal missingness rate (Table \ref{tab:missingness_proportion}), while including other $M$-nodes as covariates allowed us to control joint missingness across variables.

\begin{figure}[H]
    \centering
        \includegraphics[scale=0.69]{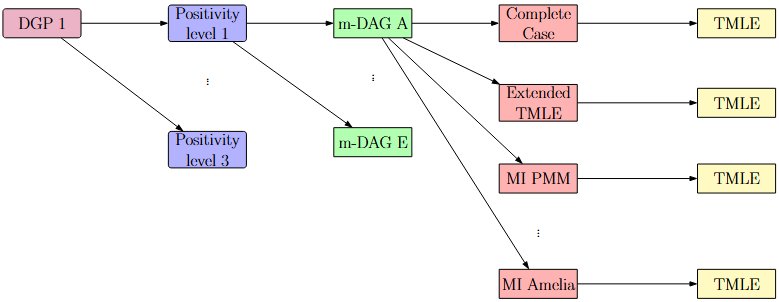}
        \caption[Flowchart]{\small Workflow illustration: The study starts with different DGPs (e.g., DGP 1), explores three positivity levels for each DGP, and incorporates m-DAGs to represent different missingness mechanisms. Finally, various missing data handling methods are applied, alongside TMLE, to estimate causal effects.}
    \label{fig:Flowchart}
\end{figure}

\subsection{Evaluation Criteria}
\label{sec:Evaluation criteria}
We assesed performance of the approaches for handling missing data, following Morris et al.\cite{refSimuGuidance}. For $n_{\text{sim}}$ simulation runs with true parameter $\theta$, the relative bias in percent was defined as
\[
\left( \frac{1}{n_{\text{sim}}}\sum_{i=1}^{n_{\text{sim}}}\hat{\theta}_i - \theta \right) \cdot \frac{1}{\theta},
\]
RMSE as
\[
\sqrt{\frac{1}{n_{\text{sim}}}\sum_{i=1}^{n_{\text{sim}}}(\hat{\theta}_i - \theta)^2},
\]
and coverage as
\[
\frac{1}{n_{\text{sim}}}\sum_{i=1}^{n_{\text{sim}}}\mathbf{1}(\hat{\theta}_{\text{low},i} \leq \theta \leq \hat{\theta}_{\text{upp},i}).
\]

\section{Design-based Simulation}
\label{sec:DGP_DesginBased}
To validate the results of the model-based simulation study, we applied the undersmoothed highly adaptive lasso (HAL) approach following Li et al. \cite{refUnderHAL}. The design-based simulation uses data from the \emph{WASH Benefits Bangladesh} study \cite{refWashData}, a cluster-randomized trial of water, sanitation, and nutrition interventions. Following Li et al., we use the binary exposure (a control group and treatment group) instead of the multilevel intervention arms, which does not preserve the original cluster-randomization.
In our simulations, we therefore do not exploit the original randomization but deliberately treat the data as observational: we use undersmoothed HAL to flexibly model the conditional distributions of the exposure and outcome. By capturing even weak dependencies between baseline covariates, treatment, and outcome, HAL induces a realistic confounding structure requiring adjustment. The outcome measure used here was a height-to-age z-score for children. We then estimate the average treatment effect of ``treatment vs.\ no treatment'' using various missing data methods in conjunction with TMLE as analysis method, adjusting for the same pre-specified prognostic baseline covariates as in Li et al. \cite{refUnderHAL} 
and the original study protocol \cite{refWashData}.

\subsection{Undersmoothed Highly Adaptive Lasso}

Highly adaptive lasso (HAL) is a nonparametric regression estimator designed to estimate complex functional parameters with minimal assumptions about the underlying data distribution. HAL constructs a linear combination of indicator basis functions to approximate the true functional parameter (i.e., the distribution of the outcome Y) and is represented as \cite{refHAL}:
\[
\psi_n = \beta_0 + \sum_{s \subset \{1,2,\dots,p\}} \sum_{i=1}^{n} \beta_{s,i} \phi_{s,i} \quad \text{with constraint} \quad \beta_0 + \sum_{s \subset \{1,2,\dots,p\}}\sum_{i=1}^{n} |\beta_{s,i}| < C
\]
where \( n \) is the sample size, \( p \) is the number of covariates, \( s \) denotes any subset of \(\{1, 2, \dots, p\}\), and \(\phi_{s,i}\) are indicator basis functions based on the covariates. Consider a simple example with \(p=2\) covariates, 
\(W_1 \in \mathbb{R}\) and \(W_2 \in \{0,1\}\). We can create 
univariate basis functions for \(s = \{1\}\) by placing an indicator at each 
distinct observed value \(\tilde{w}_{1,i}\) of \(W_1\):
\(
\phi_{1,i}(w) 
\;=\; \mathbb{I}\bigl\{W_1 \,\ge\, \tilde{w}_{1,i}\bigr\}
\) for \(i=1,\dots,n.\)
The basis function that mimics an interaction with \(s = \{1,2\}\) looks like:
\(
\phi_{1,2,i}(w) 
\;=\; \mathbb{I}\bigl\{W_1 \,\ge\, \tilde{w}_{1,i},\,W_2 \,\ge\, 1\bigr\},
\) for \(i=1,\dots,n.\) The constraint ensures that the \( L_1 \)-norm of the coefficient vector is bounded by some C \( \geq 0 \), effectively controlling the complexity of the model. This regularization promotes sparsity, allowing HAL to select relevant basis functions adaptively while preventing overfitting \cite{refHAL2017}. HAL is particularly powerful in high-dimensional settings where interactions between covariates and non-linear relationships are expected. It has been shown to perform competitively in finite samples compared to other machine learning algorithms, owing to its ability to adaptively select relevant basis functions while imposing a sparsity constraint \cite{refHAL, refUnderHAL}.

\paragraph{Undersmoothing Procedure:}
In standard HAL, the \( L_1 \)-norm bound \( C \) is typically selected using cross-validation. However, in certain scenarios, it is beneficial to allow the HAL estimator to be more flexible by relaxing the \( L_1 \)-norm constraint. This approach, known as undersmoothing, can improve the efficiency of the estimator, particularly for estimating smooth features of the data distribution \cite{refUnderHalTheory}.
The undersmoothing process involves iteratively increasing \(C\) and refitting the HAL model until the following criterion is met for all basis functions \(\phi_{s,i}\) in the initial fit \cite{refUnderHAL}:
\[
\left| P_n(\phi_{s,i}(Y - \bar{Q}_{n,c_{cv}})) \right| \leq \frac{\sigma_{n}}{\sqrt{n}\log(n)},
\]
where \( P_n \) is the empirical average function, \( \bar{Q}_{n,c_{cv}} \) is the conditional outcome model fitted using cross-validation with the \( L_1 \)-norm bound \( C_{cv} \) and \(\sigma_{n}^2 = \text{Var}(\phi_{s,i}(Y - \bar{Q}_{n,c_{cv}}))\) represents the variance of the residuals multiplied by the basis functions. This criterion ensures that the model fits the data adequately while still allowing for flexibility in capturing complex relationships. By undersmoothing, HAL can achieve asymptotically efficient estimators for pathwise differentiable target features of the data distribution, while maintaining the convergence rate \( n^{-1/3} \) \cite{refUnderHalTheory}. Undersmoothed HAL can have benefits over other popular methods because it maintains the independence of the data simulation from the estimation process by avoiding the use of identical algorithms for both tasks \cite{refUnderHalTheory}.

\subsection{Framework}

\paragraph{DGP via undersmoothed HAL:}

The undersmoothed HAL procedure was then applied twice to the sampled, fully observed data,

Initially, observations were sampled with replacement from the original dataset, with a sample size of 2000, consistent with the model-based simulation. After sampling, the undersmoothed HAL procedure was applied twice to the sampled, fully observed data: First, to fit the model with \( A \) as the outcome and \( \boldsymbol{W} \) as covariates, and second, to fit the model with \( Y \) as the outcome and \( A \) and \( \boldsymbol{W} \) as covariates. The following steps were repeated 1,000 times for the original scenario with no positivity violation:

\begin{enumerate}
  \item Covariates \( \boldsymbol{W} \) were sampled with replacement from the observed dataset, maintaining a sample size of 2,000.
  \item The first undersmoothed HAL fit was used to predict \( P(A = 1|\boldsymbol{W}) \), and then the simulated \( A \) was drawn using a binomial distribution based on these predicted probabilities.
  \item The second undersmoothed HAL fit was used to predict \( Y \) given the sampled \( \boldsymbol{W} \) and simulated \( A \). Subsequently, \( Y \) was simulated by adding random errors drawn from \( \mathcal{N}(0, \hat{\sigma}^2) \) to the predictions, where \( \hat{\sigma}^2 \) represents the residual variance from the undersmoothed HAL fit.
\end{enumerate}


\noindent We calculated the quasi-true ATE value ($= 0.065$) by first randomly drawing a large number of observations ($N = 250000$) from the empirical distribution of $\boldsymbol{W}$ and used:
\[
    \hat{ATE} = \frac{1}{N} \sum_{i=1}^{N} \left[ \hat{E}(Y_{i} | A = 1, \boldsymbol{W}_{i}) - \hat{E}(Y_{i} | A = 0, \boldsymbol{W}_{i}) \right],
\]
where we estimate $E(Y | A = 1, \boldsymbol{W})$ and $E(Y | A = 0, \boldsymbol{W})$ using the fitted undersmooth HAL model.

\paragraph{Missingness via m-DAGs:}
Missingness was introduced within the framework of m-DAGs A, B, C, D and E, as illustrated in Figure~\ref{fig:m-DAGs}. To maintain consistency and comparability, missingness was induced only in variables that also exhibited missingness in the original study. Furthermore we induced missingness in both the exposure and outcome variables. The missing data proportions were kept similar to those in the design-based simulation study, with half of the observations containing at least one missing value: 20\% missingness in the outcome, 30\% in the exposure, and 40\% in either the exposure or the outcome. We followed the same rationale for specifying the coefficients in the missingness models as described in Section~\ref{sec:model_generatingMissingness} to ensure comparable strength and balance of the induced mechanisms. \newline

All missing data methods employed in the model-based simulation study were used. The parametric MI approach with interactions (MI Int) included only two-way interactions due to collinearity issues within the MICE framework. The analysis model was TMLE, using the same settings and SuperLearner library as in the model-based simulation. Further details regarding the missingness processes and the assessment of positivity violation are provided in Appendices~\ref{Append:DesignBasedPosVio} and~\ref{Append:DesignBasedMissingness}.

\section{Results}
\begin{figure}[H]
    \centering
        \includegraphics[scale=0.73]{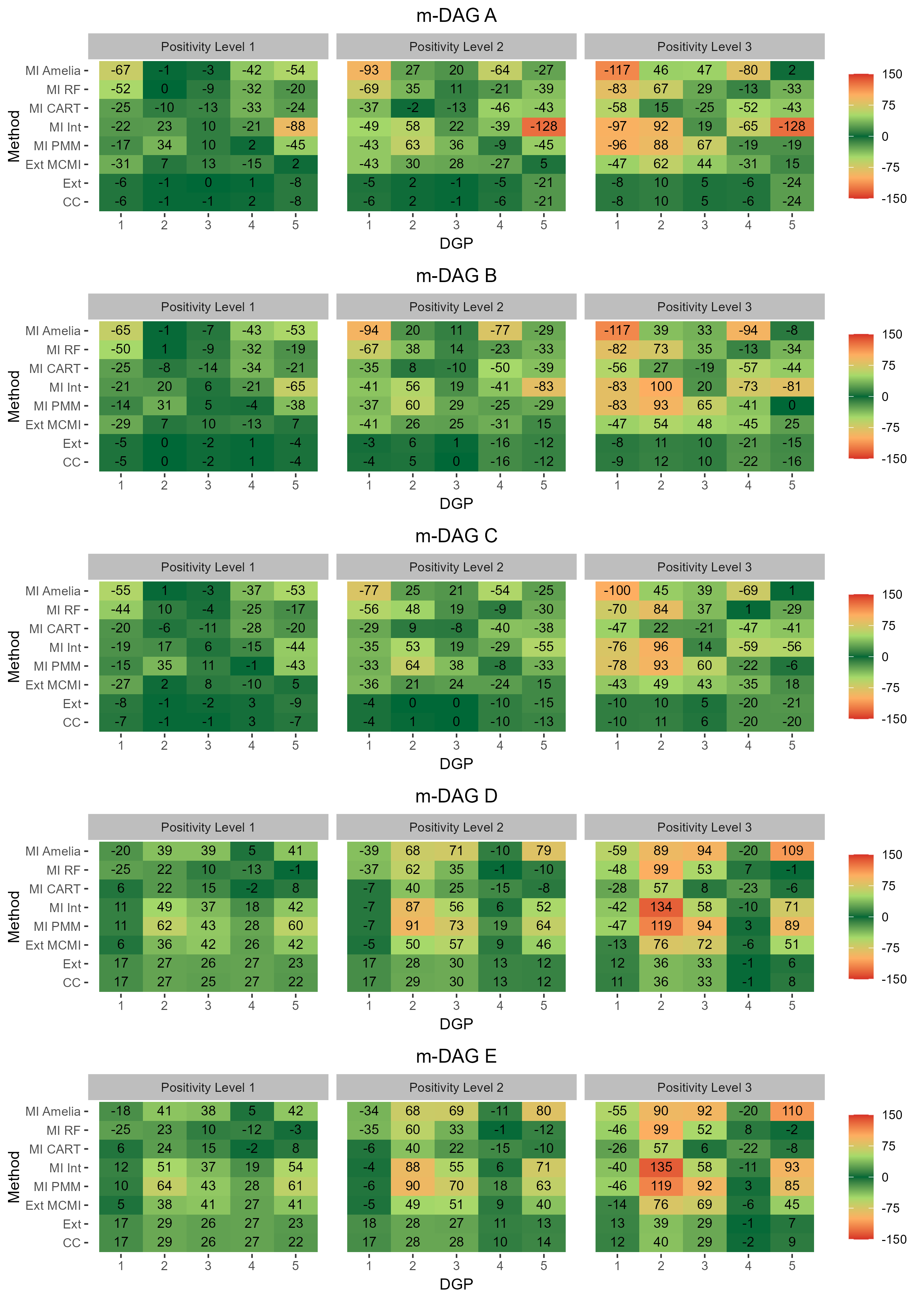}
        \caption[Relative Bias Overview for model-based Simulation]{Model-based simulation: Relative bias (\%) in ATE estimation using different missing data methods combined with TMLE and SuperLearner across the five missingness directed acyclic graphs (m-DAGs). The left, middle, and right facets display results under increasing levels of positivity violations (Level 1 $\leq 1\%$; Level 2 $\sim 10\%$; Level 3 $\sim 30\%$.), respectively.}
    \label{fig:Bias model-based Simulation}
\end{figure}

The results of the simulation studies are in Figures \ref{fig:Bias model-based Simulation}, \ref{fig:Coverage model-based Simultion}, \ref{fig:RMSE model-based Simulation}, and \ref{fig:DesignBasedWashAll}. Across simulations, three design features drive performance: (i) \emph{recoverability} of the ATE by m-DAG (A–C recoverable under no positivity violation; D–E non-recoverable), (ii) the \emph{degree of positivity violation} (Levels 1–3), and (iii) \emph{DGP complexity} (DGPs 1–5, reflecting increasing heterogeneity in variable types). Where recoverability holds and positivity is adequate, CC and Ext are expected to be approximately unbiased; with bias increasing as positivity worsens. MI methods depend on \emph{congeniality} (e.g., inclusion of interactions) and on how much imputation extrapolates beyond observed support. We therefore report patterns by (a) recoverable vs. non-recoverable m-DAGs, (b) positivity severity, and (c) DGP complexity.

\paragraph{Relative Bias in model-based simulation:}
Both CC and Ext performed similarly regarding bias across all m-DAGs, DGPs and scenarios (see Figure~\ref{fig:Bias model-based Simulation}). Specifically, in settings, where ATE was recoverable (m-DAGs A, B, and C under low positivity violation) and unbiased estimation with these two methods could be expected,  the relative bias remained low (below 10\%). With increasing positivity violation, bias also increased; however when violation was moderate (level 2), the corresponding increase in bias was likewise moderate.  At the highest level of positivity violation, bias became more pronounced, particularly in the most complex DGP, reaching up to $24\%$. In contrast, when the ATE was not recoverable under CC or Ext, both methods exhibited substantially larger bias larger bias across all DGPs compared to the recoverable m-DAGs. Notably, in the sceanrio with the highest positivity violation for DGPs 4 and 5, the ATE estimates were less biased (ranging from $1\%$ to $14\%$), despite the non-recoverability.

The third non-MI method, Ext MCMI, might yield a biased ATE estimation in all considered m-DAGs. In line with the theoretical expectations, Ext MCMI performed almost worse across all m-DAGs, DGPs, and positivity scenarios compared to CC and Ext. It consistently exhibited higher levels of bias, indicating that this method may not be reliable even when the missingness mechanism is completely at random.

The parametric MI approaches MI PMM and MI Int, often resulted in greater bias across all m-DAGs and DGPs compared to the non-MI methods CC and Ext. This was especially evident, where positivity violations were more severe (2 and 3). MI Int showed less bias in DGPs 2 and 3 across all m-DAGs in positivity levels 1 and 2 compared to MI PMM. For DGP 5, MI-PMM exhibited lower bias than MI-Int, possibly due to the inclusion of the additional gamma-distributed variable in the data.

The performance of MI-CART depended more strongly on the underlying DGP than on the specific missingness mechanism, indicating less sensitivity to recoverability and greater influence of data type and setup. Although bias increased with stronger positivity violations, MI-CART produced more stable estimates than MI-RF and both parametric MI approaches. This likely reflects that parametric MI relies on restrictive model assumptions and tends to extrapolate beyond the observed data support, whereas MI-RF may fail to select key variables as split criteria during imputation. MI-RF performed best in DGP 5 but showed the poorest performance in DGP 1. Overall, it was generally more biased than MI-CART across m-DAGs and DGPs, when positivity violation increase, except in DGP 4, where its performance was comparatively better.

MI Amelia exhibited small bias across m-DAGs A, B, and C in scenario 1 for DGPs 2 and 3. It also showed low bias, when ATE is not recoverable in m-DAGs D and E across all scenarios for DGP 4. Despite these instances, MI Amelia generally produced significantly biased results, with the bias increasing in scenarios with greater positivity violations.

We observed less biased ATE estimation across the MI methods in DGP 3 compared with DGP 2 in m-DAGs D and E, which is attributable to the higher correlation among covariates in DGP 3. Stronger inter-variable dependence can stabilize predictions in the imputation step as shown in Oberman et al. \cite{refGuidanceImpMeth}, thereby reducing bias even under MNAR mechanisms and partially mitigating the limitations imposed by non-recoverability.

\paragraph{Relative Bias in design-based simulation results:}
Both non-MI methods, CC and Ext, exhibited extremely low bias across m-DAGs A and B, where the ATE is recoverable (see Figure~\ref{fig:DesignBasedWashAll}). In m-DAG C, the ATE is no longer recoverable; however, low bias is still expected because the undersmoothed HAL fit for the conditional outcome distribution included only a few interaction basis splines, implying negligible effect modification. Bias became substantially more pronounced in m-DAGs D and E, as expected.
In contrast to the model-based simulation, the third non-MI method, Ext-MCMI, performed similarly to CC and Ext across all m-DAGs. In the design-based simulation, the Ext method was superior to both other non-MI methods across all m-DAGs.

The parametric MI approaches, MI-PMM and MI-Int, exhibited slightly greater bias across the m-DAGs where the ATE was recoverable, compared with the non-MI methods (CC and Ext). As expected, bias increased further in m-DAGs D and E.
Both methods performed similarly, as we were unable to incorporate all three-way confounder interaction effects due to collinearity issues. As reported in \cite{refAppropriateInteractions}, the appropriate inclusion of interaction effects is crucial for valid inference; thus, only a modest reduction in bias would be expected under ideal implementation.

The non-parametric MI methods (MI RF and MI CART) performed similarly across the missingness scenarios. Both exhibited higher bias in m-DAGs A, B, and C but showed markedly lower bias under m-DAGs D and E. However, MI CART was substantially superior to MI RF and yielded approximately unbiased estimates for the ATE in m-DAGs D and E.

MI Amelia also showed smaller bias in m-DAGs D and E and larger bias in m-DAGs A, B, and C. This pattern likely arises because, in D and E, the available data may more closely resemble a multivariate normal distribution than in m-DAGs A, B, and C.

Overall, the design-based simulation confirmed the theoretical expectations and was consistent with the model-based results, indicating that the ATE can be estimated with approximately negligible bias under m-DAGs A, B, and C.

\paragraph{RMSE results:}

The non-parametric MI methods (MI RF and MI CART) generally exhibited the lowest RMSE in both the model-based (Figure~\ref{fig:RMSE model-based Simulation}) and design-based (Figure~\ref{fig:DesignBasedWashAll}) simulations across all settings. In contrast, MI Amelia achieved comparable RMSE in the design-based simulation and showed lower RMSE in DGPs 2 and 3 of the model-based simulation, largely due to its lower bias (see Figure~\ref{fig:Bias model-based Simulation}). The parametric MI methods MI PMM and MI Int showed substantially higher RMSE in positvity level 2 and 3 of the model-based simulation due to high bias and performed comparably to the non-parametric methods in the design-based setting, although still slightly worse. MI Int generally exhibited higher RMSE than MI PMM.
Among the non-MI methods, Ext and CC had similar RMSE across both simulation types, while Ext-MCMI matched their performance in the design-based simulation but performed slightly better in the model-based simulation under severe positivity violations.

\paragraph{Coverage results:}

Given the number of simulation replicates ($n_{\text{sim}} = 1000$), the estimated coverage for a nominal 95\% rate is expected to fall within the range of approximately 93.6\% to 96.4\% due to Monte Carlo uncertainty \cite{refPerformenceCVTMLE}.

In both the model- and design-based simulations, MI CART and MI RF exhibited overcoverage, with MI CART generally being closer to the nominal 95\% coverage rate (see Figures~\ref{fig:Coverage model-based Simultion} and \ref{fig:DesignBasedWashAll}).
MI Amelia also showed overcoverage in most settings of both simulation types. However, in a few scenarios it produced undercoverage, and in DGP1 of the model-based simulation the undercoverage was substantial. Similarly, the parametric MI approach MI Int resulted in overcoverage for m-DAGs A, B, and C in the but suffered from marked undercoverage in m-DAGs D and E in the model-based simulation. In the design-based simulation, MI-Int achieved coverage rates closer to the nominal 95\%, though still somewhat inaccurate. It nevertheless remained closer to nominal levels than MI PMM in the model-based simulation, which is consistent with previous findings that the appropriate inclusion of interaction terms is necessary for improved inference \cite{refAppropriateInteractions}. The non-MI methods systematically led to undercoverage in both simulation types, with pronounced undercoverage in the model-based simulation under severe positivity violations (level 3) across all m-DAGs.

\section{Discussion}

Consistent with prior findings \cite{refBetancur, refDashti, refZhang2024}, CC and Ext performed as expected in settings where ATE was recoverable under these methods (m-DAGs A, B, and C) with low to moderate positivity violations, exhibiting low to moderate bias. In our simulations, positivity violations were induced at the level of the hypothetical complete data, thereby directly violating the causal identification conditions for the ATE \cite{TargetedLearningBook, refTmleNearPos, RefTmleBasicRose}. Thus, in scenarios with moderate to severe violations (levels 2–3), the ATE is non-recoverable from the observed data and biased estimates are expected.
In the model-based simulations, Ext MCMI performed significantly worse than both CC and Ext, across all m-DAGs, due to violations of recoverability conditions. Although Ext MCMI was not expected to yield unbiased ATE estimation under all m-DAGs considered, there are realistic settings where inclusion of missingness indicators is justified. Specifically, when missingness itself informs treatment assignment or prognosis. In such cases, $M$ acts as a genuine confounder and must be included to achieve identification \cite{refMICIwhen, refDaniel2012}. Our simulations instead reflected the more common situation where missingness arises from data capture rather than decision-making, so $M$ was not part of the minimally sufficient adjustment set. However, in the design-based simulation, its performance was comparable to that of CC and Ext. Overall, Ext, which incorporates an outcome-missingness model within TMLE, emerged as the most robust approach under the three non-MI methods for unbiased ATE estimation across both simulation types.

CC and Ext methods exhibited smaller bias than MI methods as positivity worsened. Likely because MI relies on extrapolation beyond the range of observed data, which can introduce bias. This issue is amplified when missing data distributions differ from the overall population \cite{refSeaman2018, refStefBuurenBook}. Parametric MI and joint modeling approaches are particularly susceptible, as their strict assumptions limit imputation accuracy compared to more flexible non-parametric methods \cite{refMIOverview}. Additionally, the Ext MCMI method showed increased bias under severe positivity violations, likely due to the use of missing indicator variables, which can introduce further bias when handling missing confounder data.

A main motivation for using MCMI lies in its ability to incorporate more observations than CC analysis, since a key limitation of CC is that many variables may have missing values, leaving few fully observed variables. A pragmatic extension could be a hybrid approach that imputes incomplete confounders via MI, excludes records with missing exposure, and addresses missing outcomes through inverse probability weighting within the TMLE framework (as in Ext). Such a strategy could also improve performance compared with Ext-MCMI. It may mitigate the biases observed for Ext MCMI by avoiding missing-indicator adjustments and explicitly modeling outcome missingness. Moreover, using MI with CART for covariates may yield coverage rates closer to the nominal level because of its flexibility in capturing complex relationships and its tendency to overestimate standard errors, which can offset the undercoverage often seen with non-MI methods and nonparametric TMLE \cite{refDashti}. Our study focused on widely used approaches: complete-case analyses, MI-based approaches, and the extended TMLE framework. Nonetheless, evaluating such hybrid strategies is a promising direction for future work. Systematic simulations could assess whether such hybrids offer a favorable gain.

Non-MI methods exhibited poor nominal coverage under severe positivity violations. Even when TMLE was well-specified, variance estimation remained challenging near positivity violations, consistent with findings by Petersen et al. \cite{RefPosVio2012} and Lendle et al. \cite{refTmleNearPos}. In contrast, MI methods maintained coverage rates closer to the nominal 95\% across all scenarios, likely due to their tendency to overestimate empirical standard errors when combined with TMLE, as noted by Dashti et al. \cite{refDashti}.

To avoid bias in parameter estimates, the imputation model must be at least as complex as the analysis model \cite{refMIOverview}. The MI Int method, which appropriately incorporates interaction effects \cite{refAppropriateInteractions}, is theoretically correctly specified for the underlying DGPs, unlike MI PMM. Previous studies \cite{refAppropriateInteractions, refTreeImpVs1} have shown that parametric MI models with interaction terms yield less biased estimates and better nominal coverage under MAR mechanisms compared to those without interaction terms. Our model-based simulations support these findings, demonstrating coverage rates closer to the nominal 95\% across all m-DAGs and levels of positivity violation than those obtained with MI-PMM. However, with respect to bias, MI-Int outperformed MI-PMM only in DGPs 2 and 3, and was inconsistent across the other three DGPs. Notably, in the design-based simulation, MI-Int did not outperform MI-PMM in terms of either bias or coverage.

Parametric imputation models with interaction terms face practical limitations, as specifying all necessary interactions can be infeasible due to collinearity and increased model complexity \cite{refStefBuurenBook}. Our results indicate that non-parametric MI methods, such as MI CART and MI RF, achieve superior nominal coverage and accuracy (measured by RMSE), consistent with previous research \cite{refTreeImpVs1}. When a fully specified parametric imputation model is impractical, MI CART is preferable, as it captures complex relationships, including interactions and nonlinearities, without requiring explicit model specification \cite{refDashti, refMIOverview}.

Previous studies have shown that effect estimates from MI RF tend to be more biased than those from MI CART, but these findings were based on MAR mechanisms and datasets containing only binary or continuous variables \cite{refCartDooveBuuren, refTreeImpVs1, refDashti}. Our model-based simulation replicated this result in DGP 1 across all five m-DAGs and scenarios. However, this pattern was not consistent across DGPs; in DGPs 4 and 5, MI RF produced less biased ATE estimates than MI CART. These findings suggest that the relative performance of MI methods depends on variable types and the complexity of the underlying DGP \cite{refGuidanceImpMeth}.

The MI Amelia method demonstrated strong overall performance on m-DAG A, B, and C under DGPs 2 and 3 in scenarios without positivity violations, likely due to the closer adherence to the multivariate normality assumption in these DGPs. In contrast, the method performed poorly under DGPs 1, 4, and 5, especially in scenarios with increased positivity violations.

\section{Conclusion}

We compared eight missing-data strategies combined with TMLE across five canonical m-DAGs under varying degrees of (structural) positivity violation. Although TMLE was used as analysis method, our identification results and qualitative patterns plausibly extend to other doubly robust estimators (e.g., AIPW) when paired with the same nuisance-learning strategy.

The MCAR and MAR scenarios (m-DAGs A and B) serve as theoretical benchmarks, while practical relevance lies in MNAR settings (m-DAGs C, D and E). When the ATE was recoverable and positivity violations were mild (e.g., m-DAG C), non-MI methods—particularly Extended TMLE (Ext)—yielded the smallest bias, while none of the evaluated methods were unbiased in non-recoverable settings (m-DAGs D and E). Under increasing positivity violation, CC and Ext generally exhibited less bias than MI approaches, whereas MI methods tended to be closer to the nominal coverage rate, reflecting a bias–coverage trade-off. Among MI strategies, MI CART provided the most reliable interval coverage and competitive RMSE, including in non-recoverable settings with severe violation. Overall, method choice should depend on recoverability and the severity of positivity violations, balancing bias, coverage, and precision.

\vspace{0.4cm}
\begin{tcolorbox}[title=Box 1: Key Takeaways, colframe=black, colback=white, sharp corners, center, width=0.80\textwidth]
\begin{itemize}
    \item \textbf{Unbiased estimation under MNAR:} Certain MNAR mechanisms (e.g., m-DAG C) permit unbiased ATE estimation; others (m-DAGs D–E) do not. Unbiased estimation does not always require the MAR/MCAR assumption.
    \item \textbf{Recoverability:} CC analysis and Extended TMLE (Ext) yield less biased estimates when the ATE is recoverable, but both methods suffer from undercoverage.
    \item \textbf{Positivity matters:} CC/Ext typically yield lower bias as positivity worsens, whereas MI methods more often achieve closer-to-nominal coverage.
    \item \textbf{Extended TMLE (Ext):} Lowest bias across mechanisms in recoverable settings but suffers undercoverage; prefer when bias reduction is the priority.
    \item \textbf{Non-Parametric MI Methods:} MI CART yields the most accurate confidence intervals (close to the nominal 95\% coverage rate) across all considered missingness mechanisms, even under severe positivity violations.
\end{itemize}
\end{tcolorbox}

\section{Applied Examples: Using the Guidance on Real Data}
\label{sec:applied-examples}

To demonstrate the practical implications of our recommendations, we illustrate their application in two empirical settings that differ in their missingness mechanisms and data complexity, thereby showcasing how the guidance can inform method selection in real-world analyses.

\paragraph{WASH Benefits Bangladesh \cite{refWashData}:}
  We reanalyzed data from the \emph{WASH Benefits Bangladesh} study \cite{refWashData}. Following Li et al.\ \cite{refUnderHAL}, we treated the analysis as observational, with a binary exposure (treatment vs. control) and the height-for-age $z$-score at the end of follow-up as the outcome. As in \cite{refUnderHAL,refWashData}, we estimated the ATE using TMLE with data-adaptive nuisance estimation and the same prespecified adjustment set. Among \(n=4863\) children, baseline covariates exhibited low levels (e.g., ‘maternal previous births’: \(4.9\%\); ‘maternal weight’: \(0.04\%\)); exposure and outcome were fully observed. \\
  Given fully observed exposure and outcome with modest covariate missingness, complete-case with TMLE represents a reasonable default approach with a small loss of information \cite{refLittle2019, refLittle2024}, when recoverability holds and positivity is adequate. In particular, the ATE is recoverable if missingness does not depend on the outcome or unmeasured causes (canonical m-DAGs A and B), and under m-DAG C when there is no treatment-effect heterogeneity across missingness patterns. Under these mechanisms, CC with TMLE yields unbiased ATE estimation. When interval coverage, rather than bias minimization, is the primary objective, MI with CART provides a pragmatic alternative, as it generally maintained closer-to-nominal coverage in our simulations, at the potential cost of increased bias in recoverable MNAR settings \cite{refLittle2019,refLittle2024}.
  
\paragraph{CHAPAS-3 efavirenz substudy \cite{refRN3856}:} 
We consider a pharmacoepidemiologic substudy of the CHAPAS-3 trial \cite{refRN3856} described in various studies \cite{refRN3860, refRN4086}. Briefly, among children with perinatally acquired HIV, the exposure of interest is the efavirenz (EFV) plasma concentration, the outcome is viral failure, and the main confounders are weight and adherence measured through pill containers with memory function. Additional relevant baseline covariates are age, sex and CYP2B6 genotype. For illustration, we focused on the first viral-load (VL) measurement at 36 weeks. Overall, there is a relevant amount of missing data: for example, about 27\% of viral load outcomes are missing. \\
Holovchak et al. \cite{refHolovchak2024} summarized their assumed missingness mechanism using a base m-DAG ($G_{\text{main}}$). In $G_{\text{main}}$, there are no arrows from the outcome to any missingness indicators, meaning that viral-load does not induce additional missingness. 
Under this scenario, and by the same reasoning, the ATE comparing low versus high EFV concentrations on viral load is recoverable \cite{refHolovchak2024}. Therefore, CC with TMLE or Extended TMLE are expected to be consistent even though the missingness is MNAR. Positivity is also of concern, as only approximately 69–85 complete cases contributed to each conditional model (exposure and outcome), which means that EFV levels with certain covariate strata were poorly supported. Together, these features indicate—consistent with our main findings—that MI combined with TMLE is likely to yield inconsistent estimates, which is also in line with the findings reported by Holovchak et al. \cite{refHolovchak2024}. Therefore, Extended TMLE should be preferred in this setting.

\section*{Data Availability Statement}
The data used in this analysis was held by Bill \& Melinda Gates Foundation in a repository. The data from ''WASH Benefits Bangladesh'' \cite{refWashData} can be found in the GitHub repository: \url{https://github.com/HaodongL/realistic_simu}

\section*{Funding information}
The research is supported by the German Research Foundations (DFG) Heisenberg Program (grants 465412241 and
465412441).

\section*{Author contribution}
All authors have accepted responsibility for the entire content of this manuscript and consented to its submission to the journal, reviewed all the results and approved the final version of the manuscript.

\section*{Acknowledgements}
We thank Margarita Moreno Betancur and Seyedeh Ghazaleh Dashti for their valuable discussions and constructive feedback that helped improve the clarity and rigor of this work.

\section*{Conflict of interest}
Authors state no conflict of interest.

\section*{Code}
The simulation was conducted in R 4.4.1 \cite{refR} and the code is available on \url{https://github.com/CWiederkehr/MissingDataTmle}.

{
\fontsize{3pt}{4pt}\selectfont
\printbibliography
}

\appendix

\section{Results}

\begin{figure}[H]
    \centering
        \includegraphics[scale=0.73]{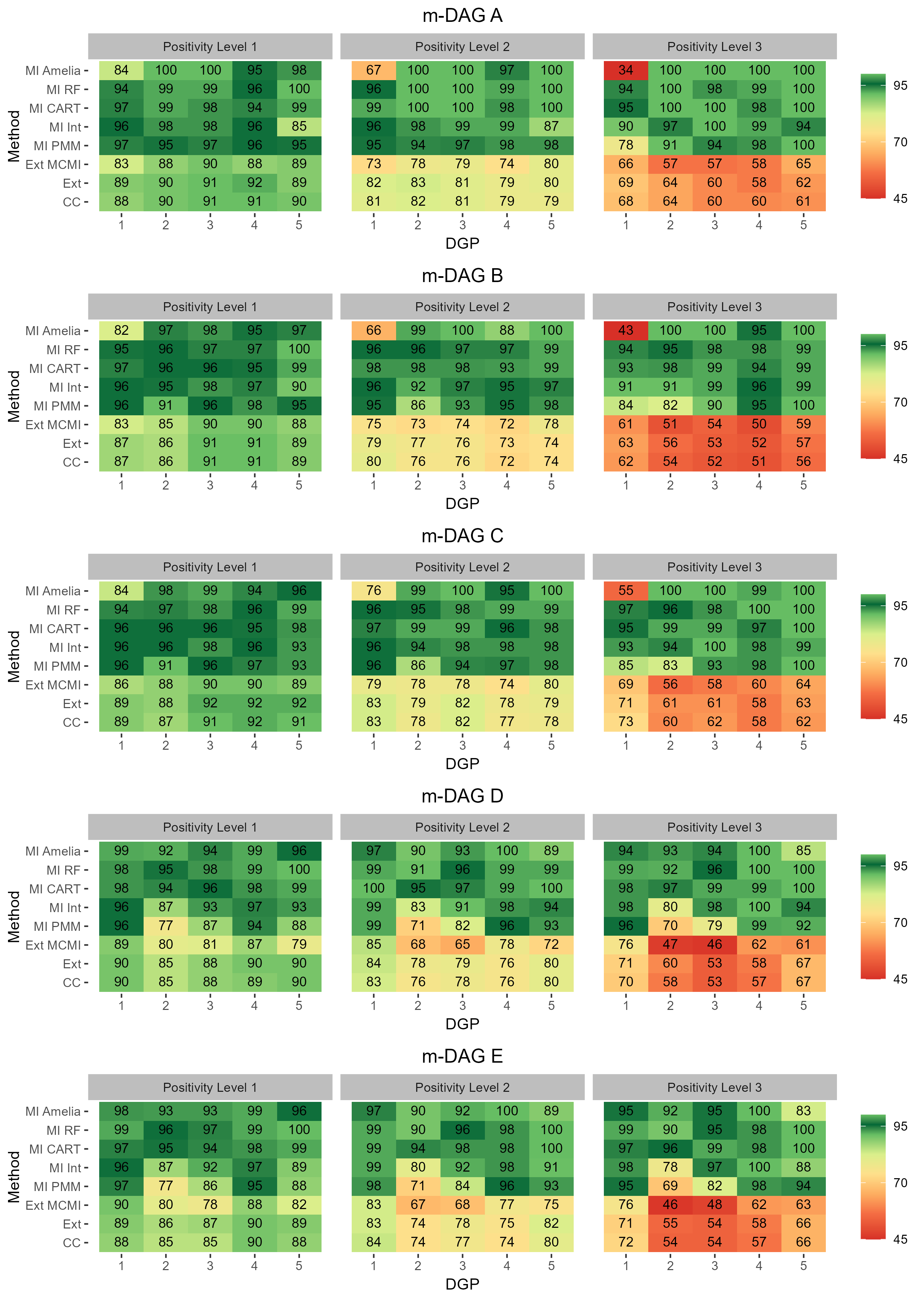}
        \caption[Coverage Overview for model-based Simulation]{Model-based simulation: Coverage in ATE estimation using different missing data methods combined with TMLE and SuperLearner across the five missingness directed acyclic graphs (m-DAGs). The left, middle, and right facets display results under increasing levels of positivity violations (Level 1 $\leq 1\%$; Level 2 $\sim 10\%$; Level 3 $\sim 30\%$.), respectively.}
    \label{fig:Coverage model-based Simultion}
\end{figure}

\begin{figure}[H]
    \centering
        \includegraphics[scale=0.75]{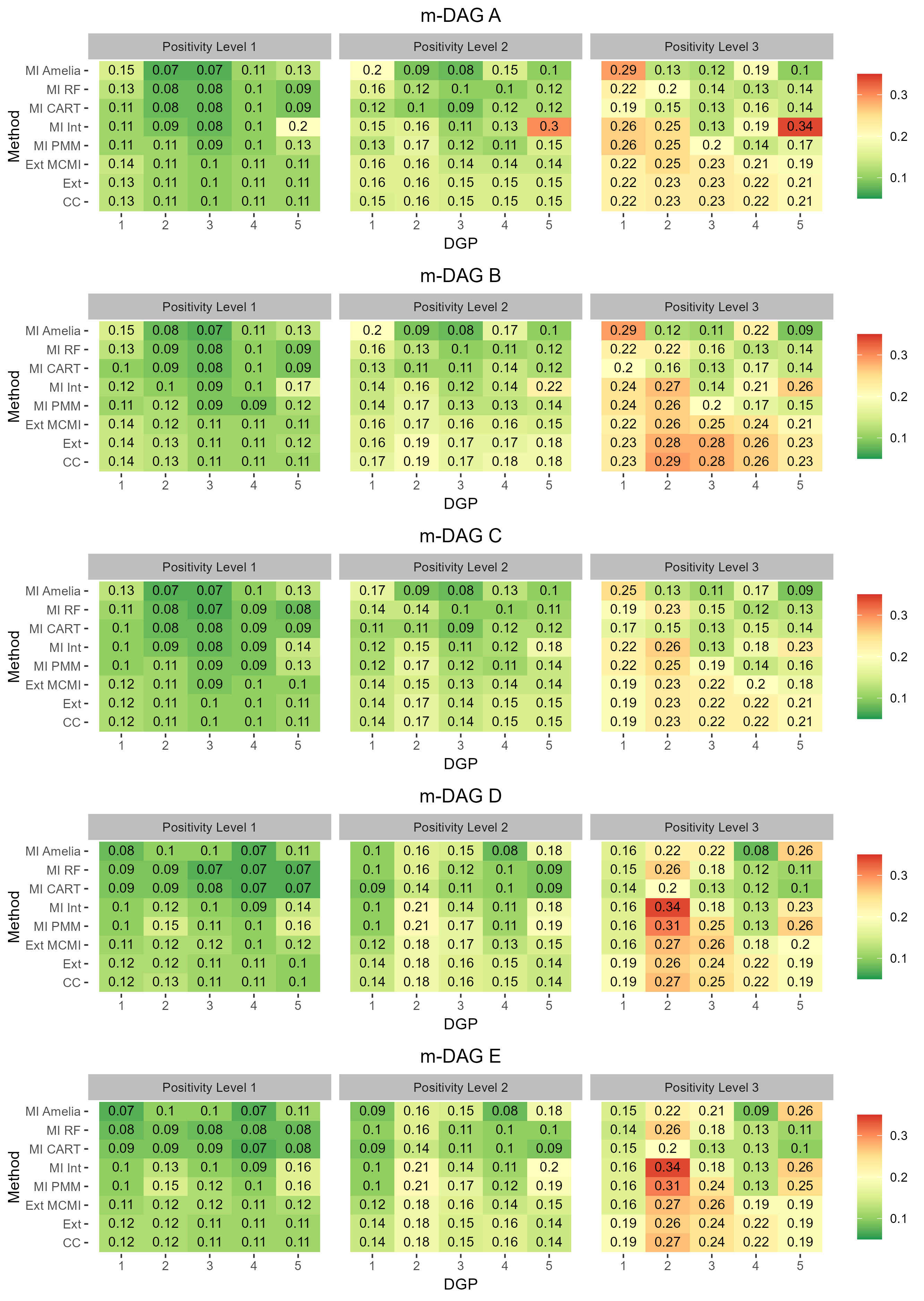}
        \caption[RMSE Overview for model-based Simulation]{Model-based simulation: RMSE in ATE estimation using different missing data methods combined with TMLE and SuperLearner across the five missingness directed acyclic graphs (m-DAGs). The left, middle, and right facets display results under increasing levels of positivity violations (Level 1 $\leq 1\%$; Level 2 $\sim 10\%$; Level 3 $\sim 30\%$.), respectively.}
    \label{fig:RMSE model-based Simulation}
\end{figure}

\begin{figure}[H]
    \centering
        \includegraphics[scale=0.78]{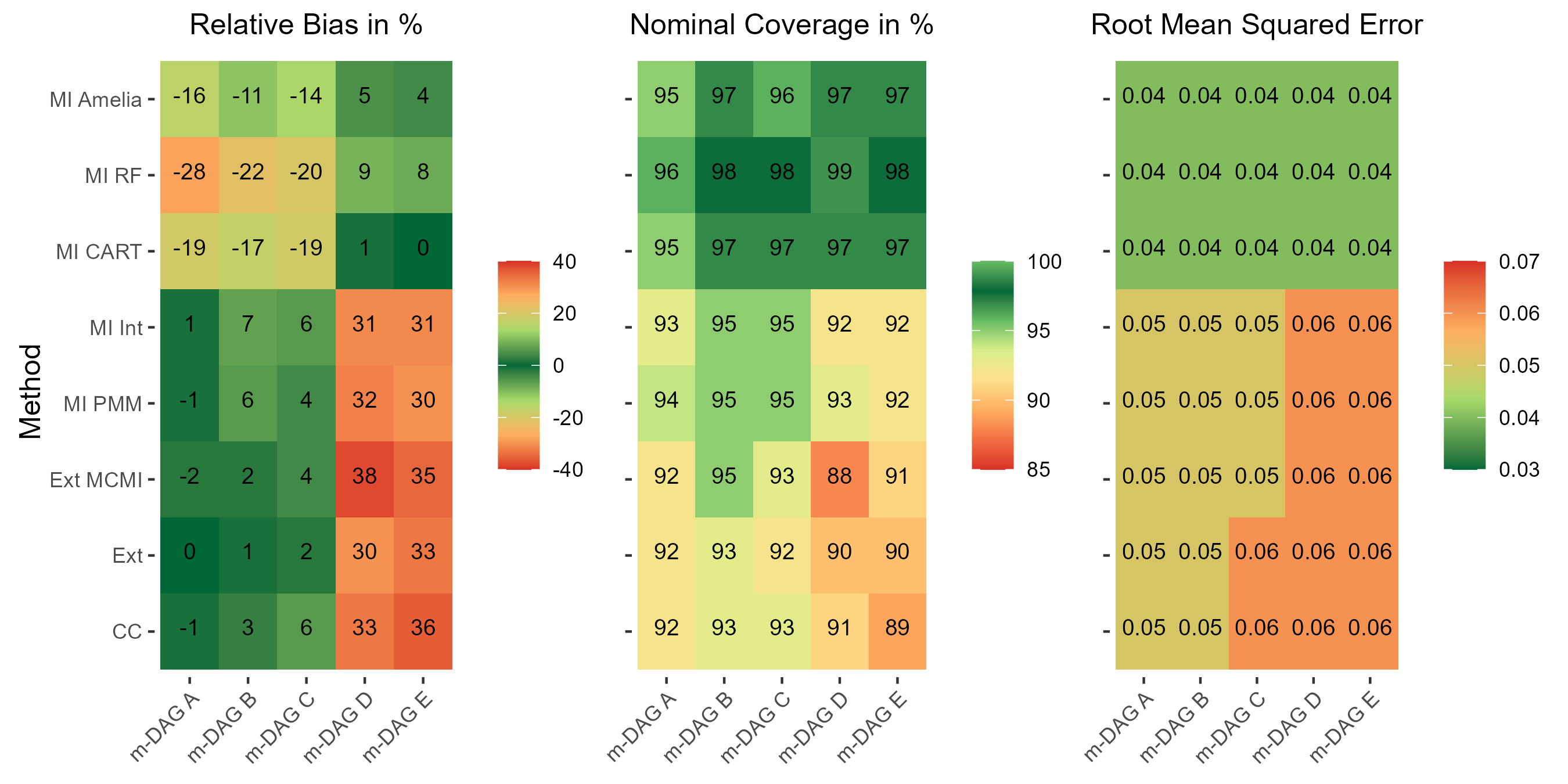}
        \caption[Result overview for design-based Simulation]{Results for the design-based simulation: Performance evaluation in ATE estimation using different missing data methods combined with TMLE and SuperLearner across the five missingness directed acyclic graphs (m-DAGs). The left panel displays the relative bias (in \%), the middle panel the nominal coverage (in \%), and the right panel the root mean squared error (RMSE).}
    \label{fig:DesignBasedWashAll}
\end{figure}

\section{Recoverability of the ATE in m-DAGs} 
\label{Appendix_Recov}

To determine whether a given m-DAG allows for unbiased estimation of the ATE, we employ the framework proposed by Moreno-Betancur  et al. \cite{refBetancur}. They describe three main conditions, required for recoverability of a target parameter, as consistency, positivity, and conditional independence conditions.
Note that \( Y \) is the continuous outcome, \( A \) is the binary exposure, \( W_{2,3,4} \) are the three partially observed confounders, and \( W_1 \) and \( W_5 \) are the fully observed confounders.

\subsection{Missingness as potential outcomes}

We define the potential outcomes in a counterfactual world without missing data. For each variable \( V \in \{Y, A, W_{2,3,4}\} \), let
\[
V^{M=0} = V^{M_Y=0, M_A=0, M_W=0},
\]
which represents the value of \( V \) when \( M_Y = M_A = M_W = 0 \), i.e., in a world without missing data in \( Y, A, W_{2,3,4} \). Here, \( M = (M_Y, M_A, M_W) \) and \( 0 = (0,0,0) \). The superscript \( M=0 \) indicates the value of the variable if, possibly contrary to fact, none of the variables were missing \cite{refBetancur}.

\subsection{Definition of Recoverability}

A target parameter \( \theta = \psi \left( P\left( Y^{M=0}, A^{M=0}, {W_{2,3,4}}^{M=0}, W_1, W_5 \right) \right) \) is said to be \textbf{recoverable} if there exists a function \( \psi \) such that
\[
\theta = \psi \left( P(O) \right),
\]
where \( P(O) \) is the distribution of the observed, incomplete data \( O \). This implies that \( \theta \) can be expressed as a characteristic of the distribution of the observed data \cite{refBetancur, refMohan2021}.

\subsection{Conditions for Recoverability}

For the target parameter to be recoverable, the following conditions must hold \cite{refBetancur}:

\begin{enumerate}
    \item \textbf{Consistency:} For individuals with observed data, the observed values coincide with their potential outcomes in the absence of missingness:
    \[
    V^{M=0} = V \quad \text{if } M_V = 0,
    \]
    for all \( V \in \{Y, A, W_{2,3,4} \} \). Additionally, the intervention that forces complete data is well-defined (e.g., missing data is not due to death).

    \item \textbf{Positivity:} For all values \( (a, y, w_1, w_5, w_{2,3,4}) \) such that
    \[
    P\left(Y^{M=0} = y, A^{M=0} = a, W_1 = w_1, W_5 = w_5, {W_{2,3,4}}^{M=0} = w_{2,3,4}\right) > 0,
    \]
    we have
    \[
    P\left(M_Y = 0, M_A = 0, M_W = 0 \mid Y^{M=0} = y, A^{M=0} = a, W_1 = w_1, W_5 = w_5, {W_{2,3,4}}^{M=0} = w_{2,3,4}\right) > 0.
    \]

    \item \textbf{Conditional Independence:} Conditional independencies in a missing Directed Acyclic Graph (DAG) are determined using d-separation, a graphical criterion that identifies whether a set of variables blocks information flow between other variables \cite{refMohan2021}. 
\end{enumerate}

\subsection{Target Parameter and Estimation via TMLE}

Our target parameter \( \theta \) is the ATE. Under the identification assumptions (see Section \ref{sec:Causal Estimand}), \( \theta \) can be expressed as follows \cite{refCausalBook}:
\[
\theta = \sum_{w} \left[ E\left(Y \mid A = 1, \boldsymbol{W} = w\right) - E\left(Y \mid A = 0, \boldsymbol{W} = w\right) \right] P(\boldsymbol{W} = w),
\]
with \( \boldsymbol{W} = (W_{2,3,4}, W_1, W_5) \) and \( w = (w_{2,3,4}, w_1, w_5) \).
To achieve a consistent and unbiased estimate of the ATE, it is necessary that both the conditional expectation \( E(Y \mid A, \boldsymbol{W}) \) and the distribution \( P(\boldsymbol{W}) \) are identifiable from the observed data. This requirement implies that both components must be expressible as functions of the observed data distribution \( P(O) \), where \( O \) denotes the observed, potentially incomplete data.

We estimate the ATE using Targeted Maximum Likelihood Estimation (TMLE):
\[
\hat{\theta}_{\text{TMLE}_n} = \frac{1}{n} \sum_{i=1}^{n} \left[ \bar{Q}_n^*\left(1, \boldsymbol{W}_i\right) - \bar{Q}_n^*\left(0, \boldsymbol{W}_i\right) \right],
\]
where \( \bar{Q}_n^*(a, \boldsymbol{W}) \) is the updated estimate of \( E(Y^{M=0} \mid A^{M=0}, {W_{2,3,4}}^{M=0}, W_1, W_5) \).

Notably, recoverability does not require the propensity score $g(A \mid \boldsymbol{W})$ to be expressible: identification is achieved via the outcome-regression functional above, without invoking an inverse-probability–weighted representation. In estimation, TMLE uses $\hat g$ solely within the fluctuation step to construct clever covariates \cite{TargetedLearningBook}. This improves efficiency and imparts double robustness but is not needed for identification if $E(Y \mid A,\boldsymbol{W})$ and $P(\boldsymbol{W})$ are identifiable.

\subsection{Recoverability in m-DAGs}
We assess the recoverability of the ATE in the five considered m-DAGs by examining whether both components of the ATE formula, the conditional expectation \( E \left( Y \mid A, W_{2,3,4}, W_1, W_5 \right) \) and the distribution \( P(W_{2,3,4}) \) are recoverable from the observed data.

\vspace{0.4cm}

\noindent \textbf{m-DAG A:} \newline
\noindent Conditional Independencies:
\begin{align*}
Y &\perp (M_Y, M_A, M_W). \\
    W_{2,3,4} &\perp (M_Y, M_A, M_W).
\end{align*}

\noindent Derivation of the Conditional Distribution of \( Y \):
\begin{align*}
P\left(Y^{M=0} \mid A^{M=0}, W_{2,3,4}^{M=0}, W_1, W_5\right) &= P\left(Y^{M=0} \mid A^{M=0}, W_{2,3,4}^{M=0}, W_1, W_5, M=0\right) & (\text{by conditional independence}) \\
&= P\left(Y \mid A, W_{2,3,4}, W_1, W_5, M=0\right) & (\text{by consistency}).
\end{align*}

\noindent Derivation of \( P(W_{2,3,4}) \): \begin{align*}
    P\left(W_{2,3,4}^{M=0}\right) &= P\left(W_{2,3,4}^{M=0} \mid M_W=0\right) & (\text{since } W_{2,3,4} \perp M_W) \\
    &= P\left(W_{2,3,4} \mid M_W=0\right) & (\text{by consistency}).
\end{align*}
Both components needed for estimating the ATE are recoverable in m-DAG A. Thus, the ATE is recoverable.
\vspace{0.3cm}

\noindent \textbf{m-DAG B:} \newline
\noindent Conditional Independencies:
\begin{align*}
Y &\perp M_Y \mid W_1, W_5, \\
Y &\perp M_A \mid W_1, W_5, \\
Y &\perp M_W \mid W_1, W_5, \\
\Rightarrow Y &\perp (M_Y, M_A, M_W) \mid W_1, W_5, \\
W_{2,3,4} &\perp (M_Y, M_A, M_W). 
\end{align*}

\noindent Derivation of the Conditional Distribution of \( Y \):
\begin{align*}
P\left(Y^{M=0} \mid A^{M=0}, W_{2,3,4}^{M=0}, W_1, W_5\right) &= P\left(Y^{M=0} \mid A^{M=0}, W_{2,3,4}^{M=0}, W_1, W_5, M=0\right) & (\text{by conditional independence}) \\
&= P\left(Y \mid A, W_{2,3,4}, W_1, W_5, M=0\right) & (\text{by consistency}).
\end{align*}

\noindent Derivation of \( P(W_{2,3,4}) \): 
\newline Same derivation as in m-DAG A because of \( W_{2,3,4} \perp (M_Y, M_A, M_W).\)
Thus, the ATE is recoverable in m-DAG B.

\vspace{0.3cm}

\noindent \textbf{m-DAG C:} \newline
\noindent Conditional Independencies:
\begin{align*}
Y &\perp M_Y \mid A, W_1, W_5, W_{2,3,4}, \\
Y &\perp M_A \mid A, W_1, W_5, W_{2,3,4}, \\
Y &\perp M_W \mid A, W_1, W_5, W_{2,3,4}, \\
\Rightarrow Y &\perp (M_Y, M_A, M_W) \mid A, W_1, W_5, W_{2,3,4}.
\end{align*}

\noindent Derivation of the Conditional Distribution of \( Y \):
Again, similar to DAG A:
\begin{align*}
P\left(Y^{M=0} \mid A^{M=0}, W_{2,3,4}^{M=0}, W_1, W_5\right) &= P\left(Y \mid A, W_{2,3,4}, W_1, W_5, M=0\right).
\end{align*}
\( P(W_{2,3,4}) \) is not recoverable since it depends on \(M_W\). Therfore the ATE is not recoverable (\textit{in the design-based Simulation}).

\noindent However in the \textit{model-based simulation}, \( \beta_A \) is constant across all values of \( \boldsymbol{W} \), implying that the treatment effect is uniform across the population. When we calculate the ATE:
\[
\text{ATE} = \sum_{w} \left[ E(Y \mid A = 1, \boldsymbol{W} = w) - E(Y \mid A = 0, \boldsymbol{W} = w) \right] P(\boldsymbol{W} = w),
\]
\noindent we find that:
\[
E(Y \mid A = 1, \boldsymbol{W} = w) - E(Y \mid A = 0, \boldsymbol{W} = w) = \beta_A,  \quad \forall \, w \in \mathcal{W}.
\]Consequently, the ATE simplifies to:
\[
\text{ATE} = \sum_{w} \beta_A \cdot P(\boldsymbol{W} = w) = \beta_A.
\]

\noindent Since the treatment effect does not vary by strata, we do not require a recoverability of \( P(W_1, W_5, W_{2,3,4}) \) for the ATE to be recoverable.
\vspace{0.3cm}

\noindent \textbf{m-DAGs D \& E:} \newline
For m-DAGs D and E the conditional expectation \( E \left( Y \mid A, W_{2,3,4}, W_1, W_5 \right) \) and the distribution \( P(W_{2,3,4}) \) are non-recoverable. The presence of collider structures such as \( A^{M=0} \rightarrow M_A \leftarrow Y \) in m-DAG D creates dependencies between \( Y \) and \( A \) when conditioning on \( M_A \). In m-DAG E, \( Y \) directly influences its own missingness indicator \( M_Y \), again violating the assumptions needed for recoverability. 

\subsection{Unbiased estimation for different missing data methods}
\label{Append:Unbiased estimation for different missing data methods}
Although the ATE is recoverable in m-DAGs A and B (and m-DAG C in the model-based simulation), when the causal identification assumptions are met. This does not guarantee that all methods combined with TMLE will yield unbiased estimates. The validity of each method depends on additional assumptions specific to its approach. Below, we discuss the applicability and limitations of various missing data methods within the context of these m-DAGs.

\begin{itemize}
   \item \textbf{Complete Case (CC) Analysis:} \\
   CC analysis includes only observations with fully observed data (i.e., \( M = 0 \), where \( M_Y = 0, M_A = 0, M_W = 0 \)). In m-DAGs A and B (and m-DAG C in the model-based simulation), CC combined with TMLE can yield unbiased ATE estimates. This holds because, as shown in the recoverability
expressions derived above, the ATE is equal in the full and complete-case data.

    \item \textbf{Extended TMLE (Ext TMLE):} \\
    Extended TMLE adjusts for potential biases due to missingness by incorporating models for the missing data mechanisms into the estimation procedure. The initial estimate \( \hat{Q}(A, \boldsymbol{W}) \) is fitted using only complete cases, the clever covariate is modified by including both the propensity score and a model for the missingness indicators conditioned on confounders and exposure. The clever covariate is defined as:
    \[
    H(1, \boldsymbol{W}) = \frac{A}{\hat{g}(1 \mid \boldsymbol{W})} \cdot \frac{\mathbb{I}(M_Y=0)}{\hat{P}(M_Y = 0 \mid A, \boldsymbol{W})}, \quad H(0, \boldsymbol{W}) = \frac{1 - A}{1 - \hat{g}(1 \mid \boldsymbol{W})} \cdot \frac{\mathbb{I}(M_Y=0)}{\hat{P}(M_Y = 0 \mid A, \boldsymbol{W})},
    \]
    where \( \mathbb{I}(M_Y=0) \) is the indicator function for \( Y \) being observed, \( \hat{g}(A \mid \boldsymbol{W}) \) is the estimated propensity score, and \( \hat{P}(M_Y = 0 \mid A, \boldsymbol{W}) \) is the estimated probability that \( Y \) is observed given \( A \) and \( \boldsymbol{W} \). For observations where \( Y \) is missing (i.e., \( M_Y = 1 \)), the updated estimate \( \hat{Q}^*(A, \boldsymbol{W}) \) remains equal to the initial estimate \( \hat{Q}(A, \boldsymbol{W}) \), as these observations do not contribute to the updating step. 
    Thus in addition to the CC method, \( P\left(Y \mid A, W_{2,3,4}, W_1, W_5, M_A = 0, M_{W}= 0, M_Y=1 \right) \) has to be recoverable, such that the predictions are unbiased and can be derived from \( P\left(Y^{M=0} \mid A^{M=0}, W_{2,3,4}^{M=0}, W_1, W_5\right) \). For m-DAG A because of the conditional independcies \( Y \perp (M_Y, M_A, M_W) \). For m-DAG B \( Y \perp (M_Y, M_A, M_W) \mid W_{1,5} \) and m-DAG C \( Y \perp (M_Y, M_A, M_W) \mid A, W_1, W_5, W_{2,3,4}. \)
    This method enables unbiased estimation for m-DAGs A and B (and m-Dag C in the model-based simulation).
    \item \textbf{Extended TMLE with Missing Covariate Missing Indicator (Ext MCMI):} \\
   Here we divide the partially observed confounders \(W_{2,3,4}\) into an observed part \(W_{2,3,4} = W_\text{obs}\) (all sets of observed confounder values) and a missing part \(W_{2,3,4} = W_\text{mis}\) (all sets of missing confounder values).
   The Missing Covariate Missing Indicator (MCMI) method relies on assumptions such as \cite{refMCMIPaper}:
   \begin{itemize}
       \item No unmeasured confounding within missingness patterns (mSITA): $$A \perp Y(a) \mid W_{1,5}, W_{2,3,4}, M_W \quad \text{for } a = 0, 1.$$
       \item The Conditionally Independent Treatment (CIT) assumption:
      \[ A \perp W_{\text{mis}} \mid W_{1,5}, W_{\text{obs}}, M_W. \]
       ( \( \rightarrow \) missing confounder values are conditionally independent of treatment given observed confounders.)
       \item The Conditionally Independent Outcome (CIO) assumption:
       \[ Y(a) \perp W_{\text{mis}} \mid W_{1,5}, W_{\text{obs}}, M_W \quad \text{for } a = 0, 1.\] 
       ( \( \rightarrow \) missing confounder values are conditionally independent of the potential outcomes)
   \end{itemize}

   CIT is violated in all considered m-DAGs because \(A\) was generated through \( P(A^{M=0} \mid W_{1,5}, W_{2,3,4}^{M=0},B) \) and because missing confounder values do have a direct effect on \(A\). The direct \(W_{2,3,4} \rightarrow A\) relationship persist whether you look at \(W_{\text{obs}}\) or \(W_{\text{mis}}\) \newline
   CIO is violated in all considered m-DAGs because \(Y\) was generated through \( P(Y^{M=0} \mid A^{M=0}, W_{1,5}, W_{2,3,4}^{M=0}) \) and since missing confounder values do have a direct effect on \(Y\).

   Since both the CIO and CIT assumptions are violated, the Ext MCMI aproach might yield a biased ATE estimation in all considered m-DAGs.
   
    \item \textbf{Multiple Imputation (MI):} \\
    Multiple imputation replaces missing values with multiple sets of simulated values, generating several complete datasets. Each dataset is analyzed separately, and the results are combined to account for uncertainty due to missingness. MI generally yields unbiased estimates when the imputation model is correctly specified and the missingness mechanism follows MAR \cite{refMIOverview}. The concept of MAR can be extended to the graphical perspective of v-MAR \cite{refMohan2021}, where missingness depends only on fully observed variables and not on partially observed ones. This condition is satisfied in m-DAGs A and B, where missingness depends solely on the fully observed confounders \( W_1 \) and \( W_5 \). In contrast, in m-DAG C, D and E, missingness depends on variables that may induce themselves to be missing (e.g., \( A \) and \( W_{2,3,4} \)), violating the v-MAR assumption. Thus, MI can provide unbiased estimates in m-DAGs A and B when used with appropriate imputation models and analysis methods, but it may not be unbiased in m-DAG C, D, and E.
\end{itemize}

\section{Model-based Simulation:}

\subsection{Data generating process:}
\label{Append:ModelBasedDGP}
DGP 3 to 5 model multivariate dependencies using a Gaussian copula. The workflow remained consistent across these DGPs; However, we present the model equations for DGP 5 here: 
\begin{enumerate}
    \item Generate samples from the Gaussian copula:
    \[ (U_1, U_2, U_3, U_4, U_5, U_6) \sim \mathcal{C}_{\text{Gaussian}}(\rho)\] with the correlation matrix $\rho$:
    \[
    \rho = 
    \begin{pmatrix}
    1 & 0.3 & -0.3 & 0.3 & 0.3 & -0.3 \\
    0.3 & 1 & 0.7 & 0.3 & 0.3 & 0.3 \\
    -0.3 & 0.7 & 1 & 0.3 & 0.7 & 0.3 \\
    0.3 & 0.3 & 0.3 & 1 & 0.7 & 0.3 \\
    0.3 & 0.3 & 0.7 & 0.7 & 1 & -0.3 \\
    -0.3 & 0.3 & 0.3 & 0.3 & -0.3 & 1
    \end{pmatrix}
    \]
    \item Transform copula samples into the desired variables $W_1$ to $W_6$:
    \begin{itemize}
        \item \( B \sim \mathcal{N}(0, 1) \)
        \item \( W_1 = \mathbb{I}(U_1 > \text{logit}^{-1}(a_0)) \)
        \item \( W_2 = \mathbb{I}(U_2 > \text{logit}^{-1}(\beta_0 + \beta_1 B)) \)
        \item \( W_3 = \text{Categorical}(p_i) \) is a categorical variable with four categories. The probabilities of each category depend on \(B \) and are created using a softmax function:
\[ P(W_3 = i) = \frac{e^{\gamma_{0_i} + \gamma_{1_i} \cdot B}}{\sum_{j=1}^{4} e^{\gamma_{0_j} + \gamma_{1_j} \cdot B}}, \quad \text{for } i,j = 1, ..., 4. \]
        \item \( W_4 = \Phi^{-1}(U_4; \mu = \delta_{0} + \delta_{1} B, \sigma = 1) \)
        \item \( W_5 = \Phi^{-1}(U_5; \mu = \zeta_{0}, \sigma = 2) \)
        \item \( W_6 = \Gamma^{-1}(U_6; \kappa, \lambda) \) is a gamma distributed variable with shape parameter: \( \kappa = 2 + \sigma_1 \cdot t_B \) and rate parameter: \( \lambda = 1 + \rho_1 \cdot t_B \) and B is symmetrical truncated such that both parameters are positive: \( t_B = \max(\min(B, 0.99), -0.99) \)
    \end{itemize}
    \item Generate exposure \( A \) and outcome \( Y \):
    Exposure \( \mathrm{A} \) was modeled via regression on \( \mathrm{B} \) and \( \mathrm{W} \) incorporating two-way confounder-confounder interactions and the outcome \( Y \) was generated through regression on \( \mathrm{A}, \mathrm{W} \) involving two-, three-, and four-way confounder-confounder interactions:
\begin{align*}
    \mathrm{A} &\sim \operatorname{Binomial}\left( 1, p \right), \\
    p &= \operatorname{logit}^{-1}(
        \eta_{0} 
        + \eta_{1} \mathrm{W}_1 
        + \eta_{2} \mathrm{W}_2 
        + \eta_{3} \mathrm{W}_3 
        + \eta_{4} \mathrm{W}_4 
        + \eta_{5} \mathrm{W}_5 
        + \eta_{6} \mathrm{W}_6 
        + \eta_{7} \mathrm{B} 
        + \eta_{10} \mathrm{W}_1 \mathrm{W}_3
        + \eta_{13} \mathrm{W}_1 \mathrm{W}_4 \\
        &\quad  
        + \eta_{14} \mathrm{W}_1 \mathrm{W}_5 
        + \eta_{15} \mathrm{W}_3 \mathrm{W}_4  
        + \eta_{18} \mathrm{W}_3 \mathrm{W}_5  
        + \eta_{21} \mathrm{W}_4 \mathrm{W}_5 
        + \eta_{22} \mathrm{W}_1 \mathrm{W}_6 
        + \eta_{23} \mathrm{W}_4 \mathrm{W}_6 
        + \eta_{24} \mathrm{W}_5 \mathrm{W}_6 
    ), \\
    \mathrm{Y} &\sim \mathcal{N}\left( \mu, \sigma = 1 \right), \\
    \mu &= 
        \upsilon_{0} 
        + \upsilon_{1} \mathrm{A} 
        + \upsilon_{2} \mathrm{W}_1 
        + \upsilon_{3} \mathrm{W}_2 
        + \upsilon_{4} \mathrm{W}_3 
        + \upsilon_{5} \mathrm{W}_4 
        + \upsilon_{6} \mathrm{W}_5 
        + \upsilon_{7} \mathrm{W}_6 
        + \upsilon_{8} \mathrm{W}_1 \mathrm{W}_3
        + \upsilon_{9} \mathrm{W}_1 \mathrm{W}_4 \\
        &\quad 
        + \upsilon_{10} \mathrm{W}_1 \mathrm{W}_5 
        + \upsilon_{11} \mathrm{W}_3 \mathrm{W}_4 
        + \upsilon_{12} \mathrm{W}_3 \mathrm{W}_5  
        + \upsilon_{13} \mathrm{W}_4 \mathrm{W}_5 
        + \upsilon_{14} \mathrm{W}_1 \mathrm{W}_6 
        + \upsilon_{15} \mathrm{W}_4 \mathrm{W}_6 
        + \upsilon_{16} \mathrm{W}_5 \mathrm{W}_6 \\
        &\quad 
        + \upsilon_{17} \mathrm{W}_1 \mathrm{W}_4 \mathrm{W}_6 
        + \upsilon_{18} \mathrm{W}_1 \mathrm{W}_5 \mathrm{W}_6 
        + \upsilon_{19} \mathrm{W}_1 \mathrm{W}_4 \mathrm{W}_5 
        + \upsilon_{20} \mathrm{W}_4 \mathrm{W}_5 \mathrm{W}_6  
        + \upsilon_{21} \mathrm{W}_1 \mathrm{W}_4 \mathrm{W}_5 \mathrm{W}_6.
\end{align*}
\end{enumerate}

\begin{table}[H]
\centering
\caption{Variable Distributions across different Data Generating Processes (DGP1 - DGP5)}
\label{tab:variable_distributions}
\begin{tabular}{llccccc}
\toprule
\multicolumn{2}{l}{\textbf{Variable}} & \textbf{DGP1} & \textbf{DGP2} & \textbf{DGP3} & \textbf{DGP4} & \textbf{DGP5} \\
\cmidrule(lr){1-2} \cmidrule(lr){3-7}
 & A (Proportion) & 0.15 & 0.15 & 0.15 & 0.15 & 0.15 \\
 & Y (Mean) & 0 & 0 & 0 & 0 & 0 \\
 & Y.SD & 1.1 & 1.3 & 1.3 & 1.5 & 1.8 \\
\cmidrule(lr){1-2} \cmidrule(lr){3-7}
 & \(W_1\) (Proportion) & 0.21 & 0.4 & 0.4 & 0.4 & 0.4 \\
 & \(W_2\) (Proportion) & 0.13 & 0.3 & 0.3 & 0.3 & 0.3 \\
 & \(W_3\) (Proportion/Mean) & 0.59 & 0.59 & 0.59 & \textit{Categorical} & \textit{Categorical} \\
 & \(W_4\) (Proportion/Mean) & 0.37 & 3 & 3 & 3 & 3 \\
 & \(W_4\) SD & - & 1 & 1 & 1 & 1 \\
 & \(W_5\) (Proportion/Mean) & 0.38 & 1 & 1 & 1 & 1 \\
 & \(W_5\) SD & - & 2 & 2 & 2 & 2 \\
\cmidrule(lr){1-2} \cmidrule(lr){3-7}
 & \(W_6\) & - & - & - & - & 2 \\
 & \(W_6\) & - & - & - & - & 1 \\
\bottomrule
&  &  &  &  &  &  \\
\multirow{1}{*}{\textit{Categorical}} & Proportion &  \({W_3}_1\)  & \({W_3}_2 \) & \({W_3}_3\) & \({W_3}_4\) &  \\
 & & 0.25 & 0.46 & 0.20 & 0.09 &  \\
\bottomrule
\end{tabular}
\end{table}

\subsection{Imposing Missingness:}
\label{Append:ModelBasedMissing}
Due to the inclusion of the additional confounder \(W_6\) in DGP5, the models used to generate missingness indicators were adjusted as follows:

\begin{align*} 
\mathrm{M}_{\mathrm{W}_{2}} &\sim \operatorname{Binomial}(1, \text{logit}^{-1}(\iota_{0} + \iota_{1} \mathrm{W}_{1} + \iota_{2} \mathrm{W}_{5} + \iota_{3} \mathrm{W}_{2}+\iota_{4} A + \iota_{5} \mathrm{Y}) ) \\ 
\mathrm{M}_{\mathrm{W}_{3}} & \sim \operatorname{Binomial}(1, \operatorname{logit}^{-1} ( \kappa_{0} + \kappa_{1} \mathrm{W}_{1} + \kappa_{2} \mathrm{W}_{5} + \kappa_{3} \mathrm{W}_{3} + \kappa_{4} A + \kappa_{5} \mathrm{Y} + \kappa_{6} \mathrm{M}_{\mathrm{W}_{2}} ) ) \\ 
\mathrm{M}_{\mathrm{W}_{4}} &\sim \operatorname{Binomial} (1, \operatorname{logit}^{-1} ( \lambda_{0} + \lambda_{1} \mathrm{W}_{1} + \lambda_{2} \mathrm{W}_{5} + \lambda_{3} \mathrm{W}_{4} + \lambda_{4} A + \lambda_{5} \mathrm{Y} + \lambda_{6} \mathrm{M}_{\mathrm{W}_{2}} + \lambda_{7} \mathrm{M}_{\mathrm{W}_{3}} ) ) \\
\mathrm{M}_{\mathrm{W}_{6}} &\sim \operatorname{Binomial} (1, \operatorname{logit}^{-1} ( \nu_{0} + \nu_{1} \mathrm{W}_{1} + \nu_{2} \mathrm{W}_{5} + \nu_{3} \mathrm{W}_{6} + \nu_{4} A + \nu_{5} \mathrm{Y} + \nu_{6} \mathrm{M}_{\mathrm{W}_{2}} + \nu_{7} \mathrm{M}_{\mathrm{W}_{3}} + \nu_{8} \mathrm{M}_{\mathrm{W}_{4}} ) ) \\
\mathrm{M}_{A} &\sim \operatorname{Binomial} (1, \operatorname{logit}{ }^{-1} ( \xi_{0}+ \xi_{1} \mathrm{W}_{1} + \xi_{2} \mathrm{W}_{5}+  \xi_{3} \mathrm{W}_{2} + \xi_{4} \mathrm{W}_{3} + \xi_{5} \mathrm{W}_{4} + \xi_{6} \mathrm{W}_{6} + \xi_{7} A + \xi_{8} \mathrm{Y} \\ &\quad + \xi_{9} \mathrm{M}_{\mathrm{W}_{2}} + \xi_{10} \mathrm{M}_{\mathrm{W}_{3}} + \xi_{11} \mathrm{M}_{\mathrm{W}_{4}} + \xi_{12} \mathrm{M}_{\mathrm{W}_{4}} ) ) \\
\mathrm{M}_{\mathrm{Y}} &\sim \operatorname{Binomial} (1 , \operatorname{logit}^{-1} (\pi_{0}+\pi_{1} \mathrm{W}_{1}+\pi_{2} \mathrm{W}_{5}+ \pi_{3} \mathrm{W}_{2}+\pi_{4} \mathrm{W}_{3}+\pi_{5} \mathrm{W}_{4} +\pi_{6} \mathrm{W}_{6} +\pi_{7} A +\pi_{8} \mathrm{Y} \\ &\quad + \pi_{9} \mathrm{M}_{\mathrm{W}_{2}} + \pi_{10} \mathrm{M}_{\mathrm{W}_{3}} +\pi_{11} \mathrm{M}_{\mathrm{W}_{4}}+\pi_{12} \mathrm{M}_{6} +\pi_{13} \mathrm{M}_{A}) )
\end{align*}

For each $V\in\{W_2,W_3,W_4, W_6,A,Y\}$, $M_V \sim \text{Bernoulli}\{\text{logit}^{-1}(\alpha_V + \sum_{U\in \text{pa}(M_V)} \beta_{VU} U)\}$, where $|\beta_{VU}|=0.6$ for binary $U$, $|\beta_{VU}|=0.2$ for continuous $U$ and for $Y$, with signs as specified in section \ref{sec:model_generatingMissingness}; $\alpha_V$ is chosen so that $E[M_V]=p_V$ equals the target proportion in Table \ref{tab:missingness_proportion}.

\newpage
\section{Design-based Simulation}

\subsection{Assesment of positivity violation:}
\label{Append:DesignBasedPosVio}
To quantify the extent of positivity violations in the synthetic design-based data, similar steps as in chapter \ref{sec:DGP_ModelBased} were conducted:

\begin{enumerate}
    \item \textbf{Model Fitting:} Fit the undersmoothed HAL model with a maximum degree of two for the basis functions to the simulated dataset. This model mirrors the data-generating HAL model in terms of the number of basis functions but is constrained to a lower degree to reduce flexibility.
    
    \item \textbf{Propensity Score Estimation:} Using the fitted HAL model, estimate the propensity scores \( \hat{P}(A = 1 \mid \boldsymbol{W}) \) for each observation in the simulated data.
    
    \item \textbf{Identification of Positivity Violations:} For each observation \( i \), compute the estimated propensity score \( \hat{p}_i = \hat{P}(A = 1 \mid \boldsymbol{W}_i) \). An observation is considered to exhibit a positivity violation if:
    \[
        \min(\hat{p}_i, 1 - \hat{p}_i) < 0.01
    \]
    \item \textbf{Quantification of Violations:} Calculate the proportion of observations that satisfy the above condition.
\end{enumerate}
We adjusted the complexity of the HAL model used for assessing positivity violations to ensure consistency with the data-generating process while mitigating overestimation of positivity violations. Importantly, we maintained the same number of basis functions as the data-generating model to preserve the structural characteristics necessary for a fair assessment.

\paragraph{Practical Considerations:}
In practice, the implementation of HAL may involve a large number of basis functions, especially when the number of covariates is high. To manage computational complexity, the number of basis functions can be reduced by limiting interactions to main terms and second-order terms or by binning continuous covariates \cite{refUnderHAL}. These adjustments, which can be controlled through the \texttt{hal9001} package \cite{refHal9001, refHal9001art}, help balance the complexity of the model against computational feasibility.

\subsection{Imposing Missingness:}
\label{Append:DesignBasedMissingness}
\begin{figure}[htp]
  \centering
  \caption[m-DAGs]{\small Missingness directed acyclic graphs (m-DAGs) illustrating the five missingness scenarios considered in the design-based simulation. Figure has been adapted from Moreno-Betancur et al. \cite{refBetancur}.}
  \includegraphics[scale=0.93]{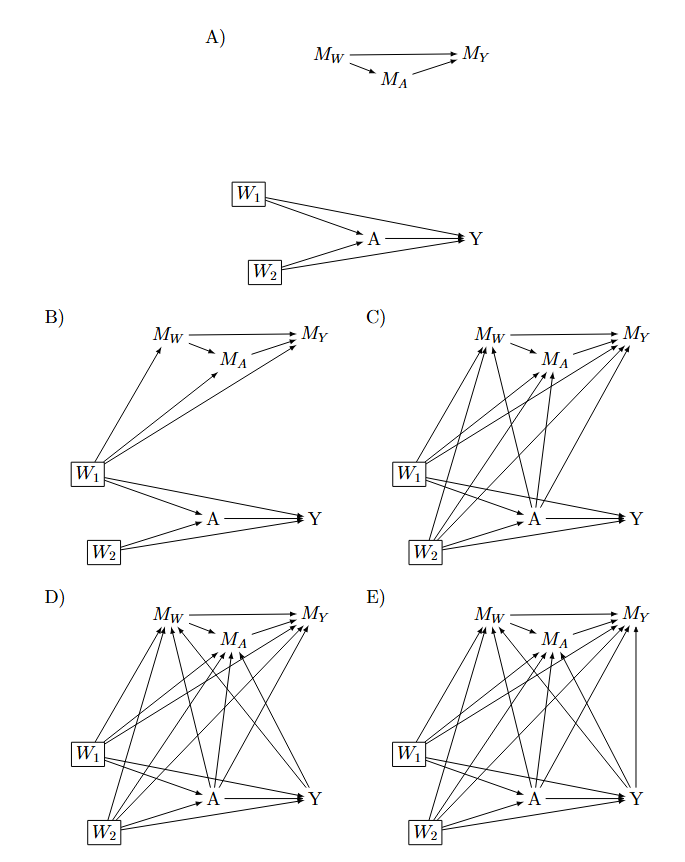} 
\caption*{\fontsize{10}{12}\selectfont For simplicity confounders without missing values $W_1$ and confounders with missing values $W_2$ are respectivley represented on single nodes. $M_W$ represents the vector of missingness indicators for all partially observed confounders at once. The m-DAG A (trivial m-DAG) represents the simplest missingness scenario and corresponds to a missing completely at random mechanism.}
\label{fig:m-DAGsHAL}
\end{figure}

Figure~\ref{fig:m-DAGsHAL} shows the same m-DAGs as for the model-based simulation. However, we did split up the confounders into a set of partially observed confounders \( W_{enwast}, W_{mhtcm}, W_{mwtkg}, W_{W_parity} \) represented through node \( W_2 \) and into a set of fully observed confounders  
\( W_{sex}, W_{month}, W_{brthmon}, W_{hfoodsec}, W_{enstunt}, \newline W_{agedays}, W_{meducyrs}, W_{nhh}, W_{mage}, W_{mbmi}, W_{feducyrs} \) represented through node \( W_1\). 

\noindent Furthermore the models used to generate missingness indicators were adjusted as follows:

\begin{align*} 
\mathrm{M}_{W_{enwast}} &\sim \operatorname{Binomial}(1, \text{logit}^{-1}(\iota_{0} + \iota_{1} \mathrm{W}_{sex} + \iota_{2} \mathrm{W}_{agedays} + \iota_{3} W_{meducyrs} +\iota_{4} W_{enwast}  +\iota_{5} A + \iota_{6} \mathrm{Y_{haz}}) ) \\ 
\mathrm{M}_{W_{mhtcm}} & \sim \operatorname{Binomial}(1, \operatorname{logit}^{-1} ( \kappa_{0} + \kappa_{1} \mathrm{W}_{sex} + \kappa_{2} W_{agedays} + \kappa_{3} \mathrm{W}_{meducyrs} + \kappa_{4} W_{mhtcm} + \kappa_{5} A \\ &\quad + \kappa_{6} \mathrm{Y}_{haz} + \kappa_{6} \mathrm{M}_{\mathrm{W}_{enwast}} ) ) \\ 
\mathrm{M}_{W_{mwtkg}} & \sim \operatorname{Binomial}(1, \operatorname{logit}^{-1} ( \lambda_{0} + \lambda_{1} \mathrm{W}_{sex} + \lambda_{2} W_{agedays} + \lambda_{3} \mathrm{W}_{meducyrs} +  + \lambda_{4} W_{mwtkg} + \lambda_{5} A \\ &\quad + \lambda_{6} \mathrm{Y}_{haz} + \lambda_{7} \mathrm{M}_{\mathrm{W}_{enwast}} + \lambda_{8} \mathrm{M}_{\mathrm{W}_{mhtcm}} ) ) \\ 
\mathrm{M}_{W_{parity}} & \sim \operatorname{Binomial}(1, \operatorname{logit}^{-1} ( \nu_{0} + \nu_{1} \mathrm{W}_{sex} + \nu_{2} W_{agedays} + \nu_{3} \mathrm{W}_{meducyrs} +  \nu_{4} W_{parity} + \nu_{5} A \\ &\quad + \nu_{6} \mathrm{Y}_{haz} + \nu_{7} \mathrm{M}_{\mathrm{W}_{enwast}} + \nu_{8} \mathrm{M}_{\mathrm{W}_{mhtcm}} + \nu_{9} \mathrm{M}_{\mathrm{W}_{mwtkg}} ) ) \\ 
\mathrm{M}_{A} &\sim \operatorname{Binomial} (1, \operatorname{logit}{ }^{-1} ( \xi_{0}+ \xi_{1} \mathrm{W}_{sex} + \xi_{2} \mathrm{W}_{agedays}+  \xi_{3} \mathrm{W}_{meducyrs} + \xi_{4} \mathrm{W}_{enwast} + \xi_{5} \mathrm{W}_{mhtcm} \\ &\quad + \xi_{6} \mathrm{W}_{mwtkg} + \xi_{7} \mathrm{W}_{parity} + \xi_{8} A + \xi_{9} \mathrm{Y_{haz}} + \xi_{10} \mathrm{M}_{\mathrm{W}_{enwast}} + \xi_{11} \mathrm{M}_{\mathrm{W}_{mhtcm}} \\ &\quad + \xi_{12} \mathrm{M}_{\mathrm{W}_{mwtkg}} + \xi_{13} \mathrm{M}_{\mathrm{W}_{parity}} ) ) \\
\mathrm{M}_{\mathrm{Y_{haz}}} &\sim \operatorname{Binomial} (1 , \operatorname{logit}^{-1} (\pi_{0}+\pi_{1} \mathrm{W}_{sex} + \pi_{2} \mathrm{W}_{agedays}+  \pi_{3} \mathrm{W}_{meducyrs} + \pi_{4} \mathrm{W}_{enwast} + \pi_{5} \mathrm{W}_{mhtcm} \\ &\quad + \pi_{6} \mathrm{W}_{mwtkg} + \pi_{7} \mathrm{W}_{parity} + \pi_{8} A + \pi_{9} \mathrm{Y_{haz}} + \pi_{10} \mathrm{M}_{\mathrm{W}_{enwast}} + \pi_{11} \mathrm{M}_{\mathrm{W}_{mhtcm}} \\ &\quad + \pi_{12} \mathrm{M}_{\mathrm{W}_{mwtkg}} + \pi_{13} \mathrm{M}_{\mathrm{W}_{parity}} +\pi_{14} \mathrm{M}_{A}) )
\end{align*}

\end{document}